\documentclass[twocolumn,showpacs,preprintnumbers,
superscriptaddress,amsmath,floatfix,prd]{revtex4}

\usepackage{graphicx}

\newcommand{\PRD}[3]{Phys.\ Rev.\ D\ {\bf #1},\ #2 (#3)}

\newcommand\g{\gamma}

\newcommand\e{\epsilon}

\newcommand\D{\Delta}

\newcommand{\non}{\nonumber\\}

\newcommand{\be}{\begin{equation}}
\newcommand{\ee}{\end{equation}}
\newcommand{\bea}{\begin{eqnarray}}
\newcommand{\eea}{\end{eqnarray}}
\newcommand{\ba}[1]{\begin{array}{#1}}
\newcommand{\ea}{\end{array}}

\begin{document}

\preprint{MIT-CTP-3712}

\title{Stressed pairing in conventional color superconductors is unavoidable} 

\author{Krishna Rajagopal}
\email{krishna@lns.mit.edu}
\affiliation{Center for Theoretical Physics, Massachusetts Institute 
of Technology, Cambridge, MA 02139}
\affiliation{Nuclear Science Division, MS 70R319,
Lawrence Berkeley National Laboratory, Berkeley, CA 94720}

\author{Andreas Schmitt}
\email{aschmitt@lns.mit.edu}
\affiliation{Center for Theoretical Physics, Massachusetts Institute 
of Technology, Cambridge, MA 02139}

\date{\today}

\begin{abstract}

At sufficiently high densities, cold dense three-flavor quark matter
is in the color-flavor locked (CFL) phase, in which all nine quarks pair
in a particularly symmetric fashion.
Once the heaviness of the strange quark (mass $m_s$) 
and the requirements of
electric and color neutrality are taken into account, 
the CFL pattern of color superconductivity requires the pairing of 
quarks that would, in the absence of pairing, have Fermi 
momenta that differ by of order $m_s^2/\mu$, with $\mu$ the 
quark number chemical potential.  
This means that at sufficiently small $\mu$, the ``stress'' on
the pairing is large enough that the system can lower its energy
by breaking pairs, resulting in some unconventional color superconductor
which includes gapless excitations, spatial inhomogeneity, counter-propagating
currents, or all three.
In this paper we ask whether there is some less symmetric but still
conventional pattern of pairing that can evade the stress. In
other words, is there a pattern of pairing in which, once
electric and color neutrality are imposed by suitable 
chemical potentials, pairing only occurs among those quarks whose
Fermi momenta would be equal in the absence of pairing? We use
graph-theoretical methods to classify 511 patterns of
conventional color superconducting pairing, and show that none
of them meet this requirement.  All feel a stress, and all
can be expected to become unstable to 
gapless modes at a density comparable
to that at which the CFL phase becomes unstable.
\end{abstract}
\pacs{12.38.Mh,24.85.+p}

\maketitle

\section{Introduction}

\subsection{Context}

At any densities that are high enough that nucleons are 
crushed into quark matter, 
the quark matter that results must, at sufficiently
low temperatures, be in 
one of a family of color superconducting phases~\cite{reviews}.
The essence of color superconductivity is quark pairing,
driven by the BCS mechanism which operates whenever there
is an attractive interaction between fermions at a Fermi 
surface~\cite{BCS}.
The QCD quark-quark interaction is strong and is attractive
between quarks that are antisymmetric in color, so we expect 
cold dense quark matter to generically exhibit color superconductivity.
Color superconducting quark matter may well occur in the
cores of compact stars.  
Except during the first few
seconds after their birth, compact stars are very cold compared
to the 10's of MeV
critical temperatures expected for color superconductivity, and
we shall therefore work at zero temperature throughout this paper.

It is by now well-established that at asymptotic densities, 
where the up, down and strange quarks can be treated
on an equal footing and the disruptive effects of the
strange quark mass can be neglected, quark matter
is in the color-flavor locked (CFL) phase, in which
quarks of all three colors and all three flavors form
Cooper pairs~\cite{Alford:1998mk,reviews}.  
The CFL phase is a color superconductor
but is an electromagnetic insulator, with zero electron density. 
To describe quark matter as may exist in the cores of 
compact stars, we need consider
quark number chemical potentials $\mu$ of order 500 MeV at most
(corresponding to baryon number chemical potentials $\mu_B=3\mu$
of order 1500 MeV at most).  The relevant range of $\mu$ is 
therefore low enough that the strange quark mass $m_s$
cannot be neglected: it is expected to be
density dependent,
lying between the current mass $\sim 100$~MeV and the
vacuum constituent quark mass $\sim 500$~MeV.  In bulk
matter, as is relevant for compact stars where we
are interested in kilometer-scale volumes, we must
furthermore require electromagnetic and color
neutrality~\cite{Iida:2000ha,Amore:2001uf,Alford:2002kj,Steiner:2002gx,Huang:2002zd} 
(possibly via mixing
of oppositely charged phases)
and allow for equilibration under the weak
interactions. All these factors work to pull apart the
Fermi momenta of the cross-species pairing that 
characterizes color-flavor locking.  At the highest
densities, we expect CFL pairing, but as the density
decreases the combination of nonzero $m_s$ and 
the constraints of neutrality put greater and
greater stress on cross-species pairing, and reduce
the excitation energies of those fermionic quasiparticles
whose excitation would serve to ease the stress by
breaking pairs~\cite{Alford:2003fq}.

If we imagine beginning in the CFL
phase at asymptotic density, reducing the density, and
suppose that CFL pairing is disrupted by the heaviness of
the strange quark before color superconducting quark matter
is superseded by the hadronic phase, the CFL phase must
be replaced by some phase of quark matter in which
there is less, and less symmetric, pairing. The nature of
this next phase down in density is currently not established.

Within a spatially homogeneous ansatz,
the next phase down
in density is the gapless CFL (gCFL)
phase~\cite{Alford:2003fq,Alford:2004hz,Alford:2004nf,Ruster:2004eg,Fukushima:2004zq,Alford:2004zr,Abuki:2004zk,Ruster:2005jc}.  
In this phase,
quarks of all three colors and all three flavors still form
Cooper pairs, but there are regions of momentum space in
which certain quarks do not succeed in pairing, and these
regions are bounded by momenta at which certain 
fermionic quasiparticles are gapless. These
gapless quasiparticles (and a nonzero density of
electrons) make the  
gCFL phase an electric conductor.
This variation on BCS pairing --- in which the 
same species of fermions that pair feature gapless 
quasiparticles --- was first proposed for two flavor
quark matter~\cite{Shovkovy:2003uu} and in an atomic
physics context~\cite{Gubankova:2003uj}.  In all these contexts,
however, the gapless paired state turns out in general to be
unstable: it can lower its energy
by the formation of counter-propagating 
currents~\cite{Huang:2004bg,Casalbuoni:2004tb}.
In the atomic physics context, the resolution of the
instability is phase separation, into macroscopic regions
of two phases in one of which standard BCS pairing occurs
and in the other of which no pairing occurs~\cite{Bedaque:2003hi}.  This 
picture appears to be supported by very recent 
experiments~\cite{KetterleImbalancedSpin,HuletPhaseSeparation}.
In the QCD context, the analogue of this phase separation would
require coexisting phases one of which is positively charged
and the other of which is negatively charged.  This may
in fact be the resolution to the instability of gapless
color superconductivity in two-flavor quark matter~\cite{Reddy:2004my}.
In three-flavor quark matter, where the instability of
the gCFL phase has been established in Refs.~\cite{Casalbuoni:2004tb},
phase coexistence would require coexisting components with opposite
color charges, in addition to opposite electric charges, making
it very unlikely that a phase separated solution can have lower
energy than the gCFL phase~\cite{Alford:2004hz,Alford:2004nf}.  
The gluon condensate phase of Ref.~\cite{Gorbar:2005rx} is another 
possible resolution to the instability in two-flavor quark
matter, but it requires an unbroken color $SU(2)$. 
In the three-flavor
quark matter context, it seems likely that a ground state with
counter-propagating currents is required.  This could take
the form of a crystalline color 
superconductor~\cite{Alford:2000ze,Bowers:2002xr,Casalbuoni:2005zp} --- the 
QCD analogue of a form of non-BCS pairing
first considered by Larkin, Ovchinnikov, Fulde and Ferrell~\cite{LOFF}.
Or, given that the CFL phase itself is likely augmented
by kaon condensation~\cite{Bedaque:2001je}, it could take the form of
a phase in which a CFL kaon condensate carries a current in one
direction balanced by a counter-propagating current in the
opposite direction carried by gapless quark 
quasiparticles~\cite{Kryjevski:2005qq}.
This meson supercurrent phase has been shown to have a lower free
energy than the gCFL phase.
The investigation of crystalline color superconductivity in
three-flavor QCD has only just begun~\cite{Casalbuoni:2005zp}, and it remains
to be seen whether such a phase can have
a lower free energy
than the meson current phase, making it a possible resolution to
the gCFL instability.  The simplest ``crystal'' structures
do not suffice, but experience in the two-flavor context~\cite{Bowers:2002xr}  
suggests
that realistic crystal structures constructed from more plane waves
will prove to be qualitatively more robust.  

\subsection{Purpose}

The analysis of unconventional pairing in color superconductivity is
challenging, given the myriad non-BCS possibilities.  In
this context, our purpose in this paper is to make an exhaustive
search through all the possibilities for conventional BCS
pairing in three-flavor quark matter that involve pairing
patterns that are less symmetric than CFL.  The first
and best studied of these is the 2SC 
phase~\cite{Bailin:1983bm,Alford:1997zt,Rapp:1997zu}, in which
only quarks of two colors and two flavors pair.  There
are, however, many other possibilities.  The question
we ask in this paper is whether any of these patterns of pairing
allow conventional BCS pairing to persist down to a value of 
the $\mu$ that is parametrically lower than that at which
the CFL phase breaks down. We shall find
that no such phases exist.

We shall only consider patterns of pairing in which
the Cooper pairs are antisymmetric in color, as that 
is the channel in which the interaction between quarks
is attractive. Furthermore, we focus on $J=0$ pairing ($J$ being the 
total angular momentum of a Cooper pair) since $J=1$ pairing between quarks of the 
same flavor yields gaps that are several orders of magnitude 
smaller~\cite{Iwasaki:1994ij,Alford:1997zt,Alford:2002rz}. 
Antisymmetry in color, together with $J=0$,
requires antisymmetry in flavor.   We shall therefore consider
all patterns of BCS pairing that are antisymmetric in both
color and flavor.

\begin{figure} [t]
\begin{center}
\includegraphics[width=9cm]{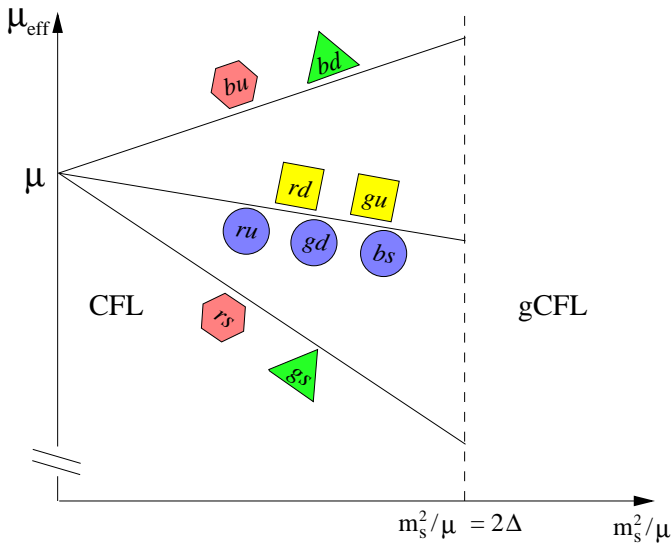}
\vspace{0.5cm}
\caption{Schematic diagram of the 
effective chemical potentials $\mu_{\rm eff}$ 
for the nine quarks as a function of the strange mass, in
the CFL phase. We abbreviate red, green and blue by $r$, $g$ and $b$
and up, down and strange by $u$, $d$ and $s$. (The effective chemical potentials are best
thought of as the Fermi momenta that the quarks would have
in the absence of pairing but
in the presence of the chemical potentials $\mu_e$,
$\mu_3$ and $\mu_8$ that serve to maintain electric and
color neutrality in the paired state.)  
The splitting between effective chemical potentials is
proportional to $m_s^2/\mu$.
The quarks encircled
with the symbol of the same shape (and color on line) pair
with each other in the CFL phase.  We see two instances
of stressed pairing --- i.e. pairing between quarks with
differing effective chemical potentials.  Such pairing
gains a free energy benefit of order $\D^2\mu^2$ but also
exacts a free energy cost of order $(m_s^2/\mu)^2\mu^2=m_s^4$.
Within a spatially homogeneous ansatz, the CFL
phase is superseded by the gCFL phase --- in which some
$bu$-$rs$ and $gs$-$bd$ pairs break --- when $m_s^2/\mu>2\D$.
However, the gCFL phase is itself unstable,   meaning that the CFL
phase must in fact be superseded earlier by some other
phase. In this paper we ask whether any conventional BCS-paired
state involves pairing only between quarks with the
same effective chemical potential.  Such a phase could
take over from the CFL phase and persist down to
a density for which $m_s^2/\mu \gg \D$.  We find no such phase.
}
\label{figsplit}
\end{center}
\end{figure}

The logic of our investigation is perhaps best explained with
reference to the CFL phase itself, see Fig.~\ref{figsplit}, although
the point of our investigation is to apply this logic to 
other phases.  In the CFL phase, color neutrality requires
a nonzero chemical potential $\mu_8$ for the color charge
proportional to $T_8$ given by $\mu_8=-m_s^2/(2\mu)$~\cite{Alford:2002kj,Steiner:2002gx}, 
whereas electric neutrality requires
$\mu_e=0$~\cite{Rajagopal:2000ff}.  
In the presence of these chemical potentials, but
in the absence of pairing, the Fermi momenta of the nine
quark species would be given by
the effective chemical potentials plotted in Fig.~\ref{figsplit}.
The merit of this figure is that one can immediately see that
the pattern of pairing requires pairing between species with
differing effective chemical potentials, separated
by $\delta\mu_{\rm eff}\sim m_s^2/\mu$.  In this circumstance,
the BCS state describing, for example, $gs$-$bd$ pairing
is constructed by first filling the
$gs$ and $bd$ states up to a common Fermi surface given by
the average between their two effective chemical potentials --- which
exacts a free energy price of 
order $\delta\mu_{\rm eff}^2 \mu^2\sim m_s^4$ --- 
and then pairing --- which gains a free energy benefit
of order $\D^2\mu^2$.    In the CFL phase at zero
temperature the gap parameter $\D$ is estimated
to be of order 10 to 100~MeV~\cite{Alford:1998mk,reviews}.
When the density drops sufficiently that the quark
chemical potential $\mu$ falls below some critical value
of order $m_s^2/\D$, then, the cost of pairing must exceed
the benefit gained from pairing,
$gs$-$bd$ pairs must break, and the gCFL phase at that point takes
over from the CFL phase. At the CFL-gCFL transition, however,
these two phases have the same free energy.  This means that
given that the gCFL phase is itself unstable, there must be
some other phase which takes over from CFL before the CFL-gCFL
transition is reached.  In this investigation, we suppose
that this next-to-highest density phase of quark matter 
involves conventional BCS pairing only, and involves
pairing that is antisymmetric in color and flavor only.
We then ask whether there is any pattern of such pairing
in which only quarks with the same effective chemical
potential pair.

If the answer to this question were yes, we would then have a candidate
pattern of pairing (or several candidate patterns) which could
persist down to much lower densities than the CFL phase,
where $m_s^2/\mu \gg \D$, without becoming unstable towards
pair-breaking, and hence without the need to face the question
of what unconventional non-BCS state
resolves the instability of a gapless color superconducting
state towards the nucleation of counter-propagating currents.  
We would then have to derive and solve the gap equation for
this stress-free 
pattern of pairing, studying it more fully than in the
analysis we present here.

In fact, we find no patterns of stress-free pairing.  Among
all the patterns of color and flavor
antisymmetric BCS pairing that are possible in three-flavor
quark matter, stressed pairing is unavoidable.
We conclude that
even if the CFL phase were supplanted by some less symmetric
pattern of otherwise still conventional BCS pairing, any such
phase would itself become unstable to pair breaking for
some $m_s^2/\mu$ of order $\D$.  This means that unless $\D/m_s^2$ 
is large enough that the 
CFL phase persists down to low enough densities that it
is superseded by hadronic matter, analysis of non-BCS
options, including spatial inhomogeneity and/or counter-propagating currents
and/or gapless color superconductivity, is unavoidable.

The remainder of the paper is organized as follows. 
In Sec.\ \ref{neutrality}, we present a
method to implement neutrality conditions in 
superconducting matter for a generic pattern of
pairing.  This  method is a generalization of that applied to
the CFL and 2SC phases in Ref.~\cite{Alford:2002kj}.  We shall assume that
$m_s/\mu$ is small, but as Fig.~\ref{figsplit} makes clear this
is sufficient to answer the question of whether stress-free
pairing is possible.
In Sec.\ \ref{toy}, we present two simple toy models with two and three fermion
species, respectively, in order to illustrate the 
interplay betweeen the pairing pattern and the neutrality 
conditions. In Sec.\ \ref{realworld} we describe the application of
our method to three-flavor quark matter, including subtleties
associated with imposing color neutrality.   Some of these
subtleties are explained further in Appendix B.  
Sec.~V contains the central argument and 
results of the paper.

\section{Neutrality conditions and stress-free states} \label{neutrality}

In this Section, we present a general form of the neutrality
conditions in a superconducting state, to leading order in
small parameters in a sense that we shall explain. In subsequent
sections, we shall use this formalism first to analyze two simple
toy models and then to analyze the multitude of pairing patterns
that are possible in quark matter with three colors and 
three flavors.  

Consider a system of $N$ fermions with a common chemical
potential $\mu$ and $K$ different 
``charge chemical potentials'' $\mu_{Q_k}$, meaning
that the $N$ fermions have chemical potentials 
\be \label{sumchem}
\mu_i = \mu + \sum_{k=1}^K q_{ik}\mu_{Q_k} \,\, , \qquad i=1,\ldots,N \,\, .
\ee
The coefficient $q_{ik}$ is the $Q_k$-charge of the $i$-th fermion.
Neutrality with respect to $K$ different 
$Q_k$-charges then requires that the free energy $\Omega$ satisfy
\be \label{neutral}
\frac{\partial\Omega}{\partial\mu_{Q_k}} = 0 \,\, , \qquad k=1,\ldots , 
K \,\,.
\ee
In Sections IV and V, where we shall
apply this formalism to quark matter, $N$ will be 9 (quarks of
three colors and three flavors) and $K$ will be 3, because
electric and color neutrality are maintained via the adjustment
of three chemical potentials $\mu_e$, $\mu_3$ and $\mu_8$ coupling
to electric, color-$T_3$, and color-$T_8$ charges.
In QCD, $\mu_e$, $\mu_3$ and $\mu_8$ are the zeroth components of
electromagnetic and color gauge fields, and the gauge field dynamics
ensure that they take on values such that the matter is 
neutral, satisfying 
(\ref{neutral})~\cite{Alford:2002kj,Gerhold:2003js,Dietrich:2003nu}.

We shall only consider charges $Q_k$
which are traceless,
\be
\label{traceless}
\sum_i q_{ik} = 0 \ ,
\ee
as this suffices for the analysis of three-flavor quark 
matter.   
If all the charges with 
respect to which neutrality must be maintained are 
traceless, as we assume, then 
if all the fermions have the same mass in
the absence of pairing the matter is neutral
with all $\mu_{Q_k}$ set to zero.
We shall assume, however, that some of the fermions have
mass $m$ while others are massless.
(In three-flavor quark matter $m$ will 
be the strange quark mass, and we shall treat the light quarks
as massless.)  We shall always assume that 
$m$ and all the $\mu_{Q_k}$ are much smaller than the common $\mu$.
We shall then see that neutrality requires that the 
$\mu_{Q_k}$ are parametrically of order $m^2/\mu$.

Our 
analysis can be applied to three-flavor quark matter in
which not all quarks pair, but it cannot be applied directly to 
two-flavor quark matter since in that context electric
charge is not traceless. However, the question that 
we address in three-flavor quark matter in this paper
can be answered with much less effort in the two-flavor case,
in which $\mu_e$
is parametrically of order $\mu$ in the absence of pairing.
This makes it easy to show that there are no 
stress-free patterns of pairing in two-flavor quark matter that
are antisymmetric in color and flavor. 

We shall consider many different patterns of pairing, but
in each case we shall simplify our notation 
by pretending that any quarks which pair
do so with a common gap parameter $\Delta$, assumed
to be $\ll\mu$.  In any realistic context, 
the pattern of pairing obtained by solving some
gap equation may well have different gap parameters
for different pairs of quarks.  In such a case,
the $\Delta$ appearing in our analysis should be taken
to be the smallest nonzero gap parameter. 
We shall also see, momentarily, that the magnitude 
of the gap parameter $\Delta$
does not actually play a role in our analysis; all that matters
to us is the pattern of which quarks pair.

The free energy of a 
conventional BCS state can be described as if one first
assigns a common Fermi momentum $\nu$ to all the fermions
that pair with each other, and then reduces the free energy
relative to that of
this fictional unpaired state by a condensation
energy of order $\Delta^2\nu^2$.  If the ``free energy benefit''
of pairing (of order $\Delta^2\nu^2$) is larger than the 
``free energy cost'' of first equalizing the Fermi momenta,
then the BCS state under consideration is stable
against pair-breaking.  In this paper, we shall seek
``stress-free states'', for which the pattern of pairing
is chosen such that the free energy cost of equalizing the
Fermi momenta is zero.

As one can see from Fig.\ \ref{figsplit}, the
CFL state has four sets of quarks that are ``connected'' by
its specific pattern of pairing, meaning that in its analysis
we will have to introduce four different common Fermi momenta $\nu_n$.
Anticipating the terminology from graph theory that we shall use in 
Sec.\ \ref{realworld}, we term these sets of fermions {\em components}. 
The components of the CFL 
pattern are $\{bu,rs\}$, $\{bd,gs\}$, $\{rd,gu\}$, and $\{ru,gd,bs\}$. 
Unpaired quark matter 
has nine components, each consisting of a single quark. 
For the general case, we denote the components by ${\cal C}_n$,
with $n=1\ldots c$ where $c$ denotes
the number of components. Thus, $c=4$ for the CFL pattern. 
We shall assume that the $n$-th component ${\cal C}_n$ 
contains $r_n$ massless fermions 
and $s_n$ fermions with mass $m$.
We now give an expression for the
free energy $\Omega$, which generalizes that in Ref.~\cite{Alford:2002kj},
assuming that fermions in components ${\cal C}_n$ pair with
gap parameter $\Delta$:
\begin{eqnarray} \label{Omega}
\Omega &=& \frac{1}{\pi^2}\sum_{n=1}^c\int_0^{\nu_n}dp \,p^2 \,
\left(r_n\, p + s_n\sqrt{p^2 + m^2}
-\sum_{i=1}^{r_n+s_n}\mu_i\right) \nonumber\\
&\quad &\qquad - {\cal N}\frac{\Delta^2\mu^2}{4\pi^2}\,\, .
\end{eqnarray}
We have denoted the common Fermi momenta by $\nu_n$; these are best
thought of as variational parameters and will be determined below
by minimizing $\Omega$ with respect to them.  The condensation
energy is written somewhat schematically, as it is
of no consequence in our analysis as we now 
explain.  As long as $\Delta\ll \mu$,
as we assume, we can neglect terms of order $\Delta^2 \mu \,\mu_{Q_k}$ and
$\Delta^2 \mu_{Q_k}^2$
in the condensation energy, as they are much smaller than
the $\mu^2 \mu_{Q_k}^2$ terms that will turn out to determine
the free energy cost of equalizing Fermi momenta at the
values $\nu_i$.  Once the $\Delta^2 \mu_{Q_k}^2$ terms are dropped,
the condensation energy is as written, with ${\cal N}$
a number that depends on the pattern of pairing. 
(${\cal N}$
counts the number of pairs of fermions that pair; ${\cal N}=12$ in
the CFL pattern, for example.)
Upon making this approximation the condensation energy 
is independent of the charge chemical potentials
and hence irrelevant in the neutrality conditions. 
We therefore neglect it henceforth.  Another way of saying
this is that our purpose is to seek patterns of pairing
that minimize the free energy cost of equalizing the Fermi
momenta of those quarks that pair.  If we can eliminate
this cost, we will have stress-free pairing.  As we will not
be evaluating the free-energy benefit of pairing, the detailed
form of the condensation energy is of no consequence.

We now minimize $\Omega$ with respect to the $c$ different
common Fermi momenta $\nu_n$, one each for the paired fermions in
each of the $c$ different components ${\cal C}_n$.
In so doing, we determine
the values of the $\nu_n$ in terms of the chemical potentials
and $m$.
Because it is clear from Eq.\ (\ref{Omega}) that the  
equations $\partial\Omega/\partial\nu_n =0$ are not coupled, 
we can minimize $\Omega$ with respect to each $\nu_n$ 
separately, yielding the condition
\be
\nu_n^2\left(r_n\,\nu_n+ s_n\,\sqrt{\nu_n^2 + m^2}
-\sum_{i=1}^{r_n+s_n}\mu_i\right)=0 
\ee
that must be satisfied by each of the $\nu_n$.
The nontrivial solution to this condition is 
\begin{eqnarray}
\nu_{n} &=& \frac{r_n}{r_n^2-s_n^2}\sum_{i=1}^{r_n+s_n}\mu_i\nonumber\\ 
&-& \frac{s_n}{r_n^2-s_n^2} 
\sqrt{\left(\sum_i^{r_n+s_n}\mu_i\right)^2 + (r_n^2-s_n^2)\,m^2} \,\, .
\end{eqnarray}
We now use Eq.\ (\ref{sumchem}), and expand this expression
for $\nu_n$ to linear
order in the charge chemical potentials $\mu_{Q_k}$ and to
quadratic order in $m$.  If, upon working at lowest nontrivial
order in these small quantities, we were to 
find a stress-free pattern of pairing, we would then have to
return to this point in the analysis and investigate higher 
order corrections. As we shall show that there are no stress-free
patterns to this order, we shall not need to consider higher
order terms.  Upon making this approximation, the result
for $\nu_n$ is most conveniently written in terms of ``effective
chemical potentials'' $\mu_i^{\rm eff}$ for each of the
fermions which, following Ref.~\cite{Alford:2002kj}, we define as
\begin{eqnarray}
\mu_i^{\rm eff} &\equiv& \mu_i \qquad\quad\,
{\rm \ for\ the\ massless\ fermions}\nonumber\\
\mu_i^{\rm eff} &\equiv& \mu_i-\frac{m^2}{2\mu} 
{\rm \ \ for\ the\ fermions\ with\ mass}\  m\ .
\label{mueffdefn}
\end{eqnarray}
We find
\be \label{common}
\nu_{n} = \frac{1}{r_n+s_n}\sum_{i=1}^{r_n+s_n} \mu_i^{\rm eff} \,\, .
\ee
We see that, to the order at which we are working, the common Fermi
momentum $\nu_n$ of the fermions in the component
${\cal C}_n$ is given by the arithmetic mean of 
the effective chemical potentials of all the fermions in 
${\cal C}_n$. 
This result generalizes the one of Ref.\ \cite{Alford:2002kj}. 
It is also straightforward, although
somewhat less instructive, to write the expression for $\nu_n$
without introducing effective chemical potentials:
\be \label{common2}
\nu_{n}=\mu + \frac{1}{r_n+s_n}\sum_{i=1}^{r_n+s_n}\sum_{k=1}^Kq_{i k}\,\mu_{Q_k} - 
\frac{s_n}{r_n+s_n}\,\frac{m^2}{2\mu} \,\, .
\ee

We now evaluate the neutrality conditions (\ref{neutral}).
The straightforward but lengthy way to do this is to perform
the $p$-integral in  Eq.\ (\ref{Omega}), insert the result 
(\ref{common2}) for $\nu_n$ into the resulting
expression, and differentiate with respect to $\mu_{Q_k}$.
Both the calculation and the form 
of the result are, however, simpler if we first make the following observation.
Consider $\Omega$ as a function of $\nu_1,\ldots,\nu_c$ 
and $\mu_{Q_1},\ldots,\mu_{Q_K}$, where 
each $\nu_n$ itself is a function 
of $\mu_{Q_1},\ldots,\mu_{Q_K}$. Then, the neutrality conditions 
can be written as
\be
0=\frac{d\Omega}{d\mu_{Q_k}}= 
\sum_{n=1}^c\frac{\partial\Omega}{\partial\nu_n}
\frac{\partial\nu_n}{\partial\mu_{Q_k}}
+\frac{\partial\Omega}{\partial\mu_{Q_k}}\,\, .
\ee
The first term on the right-hand side of this equation vanishes 
as long as we are evaluating it at values of the $\nu_n$ for
which $\partial\Omega/\partial\nu_n=0$.
Consequently, we need only evaluate the second term in which
only the explicit $\mu_{Q_k}$-dependence of $\Omega$ is relevant.
We see from (\ref{Omega})
that the $\mu_{Q_k}$ occur in $\Omega$ only in terms
proportional to $\int^{\nu_n} dp \, p^2 \sim \nu_n^3$, and
hence conclude that the neutrality conditions take the form
\be
0=\frac{\partial\Omega}{\partial\mu_{Q_k}}= 
-\frac{1}{\pi^2}\sum_{n=1}^c\frac{\nu_{n}^3}{3}
\frac{\partial}{\partial\mu_{Q_k}}\sum_{i=1}^{r_n+s_n}\mu_i\ .
\ee
We can simplify this upon noting from (\ref{sumchem}) that
\be
\frac{\partial}{\partial\mu_{Q_k}}\sum_{i=1}^{r_n+s_n}\mu_i
= \sum_{i=1}^{r_n+s_n} q_{ik}
\ee
and by rewriting the double sum (first over components, then
over fermions within components) as a single sum over all the
fermions, assigning the $i$-th fermion a $\nu_i$ given by
the $\nu_n$ of the component to which it belongs.  The resulting
neutrality conditions take the form
\be
\label{semifinal}
0=\sum_{i=1}^N \nu_i^3 q_{ik}\ ,\qquad k=1,\ldots,K \,\, ,
\ee
which is as one might have expected.
We can further simplify this result by thinking of $\nu_i$
as $\mu-(\mu-\nu_i)$, noting that our
analysis is valid only to linear order in $m^2$ and $\mu_{Q_k}$
and hence, according to (\ref{common2}), only to linear order in $(\mu-\nu_i)$,
and therefore linearizing $\nu_i^3$ as
\be
\nu_i^3 = \mu^3 - 3\mu^2(\mu-\nu_i) 
= 3\mu^2\left(\nu_i-\frac{2}{3}\mu\right)\ .
\ee
Upon using the traceless condition (\ref{traceless}), the
neutrality conditions then take on the simple form
\be \label{final}
0 = \sum_{i=1}^N \nu_{i}\,q_{ik} \,\, . 
\qquad k=1,\ldots,K \,\,, 
\ee
In Sections IV and V we shall use the expression (\ref{common2}) for
the $\nu_i$'s to
apply this condition
to a multitude of potential pairing patterns in three-flavor
quark matter, after first applying it in a toy-model context
in Section III.

With the notation that we have by now introduced,
the condition for stress-free pairing can be written simply as
\be \label{robust}
\nu_{i}=\mu_i^{\rm eff} \,\, , \qquad i=1,\ldots,N \,\, .
\ee
A stress-free state has the property
that the chemical potentials that maintain
its neutrality are such
that when we write the free energy for the BCS state
we discover that those fermions which pair (i.e.
those that lie within the same component) already
have the same $\mu^{\rm eff}$. Hence,
the
arithmetic mean in (\ref{common}) yields simply (\ref{robust}).
Our goal, then, is to find 
a neutral, stress-free, 
pairing pattern, namely one 
that simultaneously fulfills Eqs.\ (\ref{final}) and
(\ref{robust}).

\section{Two toy models} \label{toy}

In this Section, we illustrate the interplay between the neutrality 
conditions and 
the pairing pattern in
two toy models. The first model consists of two fermion
species, one of which is massive. In the second model, 
we add a third, massless, fermion to this
system in order to allow for more pairing patterns.

In the first model, we have two fermion species --- $N=2$ --- 
which means that the most general charge assignments we can consider
given the requirement of tracelessness
is simply the case with $K=1$, and charge assigments
$q_1=q$ and $q_2=-q$. That is, we have fermions with chemical potentials
\bea
\mu_1 &=& \mu + q\mu_q \,\,  , \\
\mu_2 &=& \mu - q\mu_q \,\, .
\eea
We assume the first (second) fermion to be massless (massive), 
making the effective chemical potentials
\bea
\mu_1^{\rm eff} &=& \mu_1 \,\,  , \\
\mu_2^{\rm eff} &=& \mu_2 -\frac{m^2}{2\mu} \,\, .
\eea 
There are only two possible patterns: pairing and no pairing. 
Without pairing, the common Fermi 
momenta are simply the effective chemical potentials, 
$\nu_{i}=\mu_i^{\rm eff}$ and the neutrality condition 
(\ref{final}) yields
\be
\label{toymuqneutral}
\mu_q = -\frac{m^2}{4q\mu}
\ee
This is easy to understand: with $m=0$, 
the system is neutral with $\mu_q=0$. Switching on the mass 
reduces the effective Fermi surface of the second 
species reducing its density. Consequently, a nonzero $\mu_q$ is
required in order to restore neutrality, which here
simply means equality of the two Fermi surfaces. The value of
$\mu_q$ is such that
$\mu_1^{\rm eff} = \mu_2^{\rm eff} = \mu-m^2/(4\mu)$.

When there is Cooper pairing between the two species, 
there is a single common Fermi momentum
\be
\nu_{1} = \nu_{2} = \frac{\mu^{\rm eff}_1 + \mu_2^{\rm eff}}{2} = 
\mu - \frac{m^2}{4\mu} \,\, .
\ee
In this case, the neutrality condition (\ref{final}), and
indeed even the result (\ref{semifinal}), are automatically
satisfied, for any value of $\mu_q$. In this sense,
this toy model is too simple, but let us continue
with it a little farther before generalizing it.
Since we can achieve neutrality with any $\mu_q$, 
let us choose it according to (\ref{toymuqneutral}), which
makes the effective potentials for the two fermions equal
before pairing, satisfying (\ref{robust}) and rendering 
the paired state stress-free. The fact that $\mu_q$ can
be varied around its stress-free value without upsetting
neutrality is a sign that the material is an insulator.
Varying $\mu_q$ away from that in (\ref{robust})
introduces a stress on the pairing.  There is then a free-energy
cost to equalizing the Fermi surfaces, and the paired state
is only favored if its condensation energy is large enough.
The paired state will therefore be favored in a band
of values of $\mu_q$ centered on the stress-free
value (\ref{robust}), with the width of this band being
controlled by the magnitude of $\Delta$.  In our analysis,
we are not evaluating the condensation energy and so cannot
evaluate the width of the bandgap of this insulating state.

We can augment the toy model slightly by introducing ``electrons'',
massless fermions which by hypothesis do not pair and which
have charge $q$ and chemical potential $\mu_q\ll\mu$.  These
add a term of order $\mu_q^3$ to the neutrality condition, which
is subleading relative to those we have already dropped in writing
(\ref{final}), and so if we were using (\ref{final}) we would
not be able to evaluate the consequences of adding electrons.
However, as we have satisfied (\ref{semifinal}), in this simple
model, we are justified in including electrons in the neutrality
condition, and thus concluding that once we have introduced
electrons into the theory we must set $\mu_q=0$, keeping the
electron density zero, in order to maintain neutrality.
We now see that in this toy model augmented with the 
addition of ``electrons'', the stress-free pairing condition $\mu_q=-m^2/(4q\mu)$
and neutrality condition $\mu_q=0$ cannot both be satisfied.
Hence, pairing in neutral matter will of necessity be stressed,
and the BCS state will only be favored if $\Delta$ is
large enough or, equivalently, if for some fixed $\Delta$
the stress  $| \mu_1^{\rm eff}-\mu_2^{\rm eff} |$
is small enough.
In fact, in this simple context the quantitative criterion
is easily evaluated: breaking a pair costs $2\Delta$ in condensation
energy and benefits $
| \mu_1^{\rm eff}-\mu_2^{\rm eff} | = m^2/(2\mu)$ in reduced stress, and
so the paired state is favored as long as $\Delta > m^2/(4\mu)$.
This toy model thus proves in many ways to be analogous to the
CFL phase, analyzed in 
Refs.~\cite{Rajagopal:2000ff,Alford:2003fq,Alford:2004hz}.

We now consider a second toy model, in which there are several
possible pairing patterns to consider.  We add a third fermion 
species, so now $N=3$, but we continue to keep $K=1$.  We choose
one of the three fermions to be massive, and hence consider
a toy model in which the effective chemical potentials are
\begin{subequations}
\label{chempot}
\bea
\mu_1^{\rm eff} &=& \mu + q_1\mu_q \,\, , \\
\mu_2^{\rm eff} &=& \mu + q_2\mu_q \,\, , \\
\mu_3^{\rm eff} &=& \mu + q_3\mu_q  - \frac{m^2}{2\mu} \,\, ,
\eea
\end{subequations}
where
\be
q_1 + q_2 + q_3 = 0 \,\, .
\label{condition}
\ee

\begin{figure} [t]
\begin{center}
\includegraphics[width=8cm]{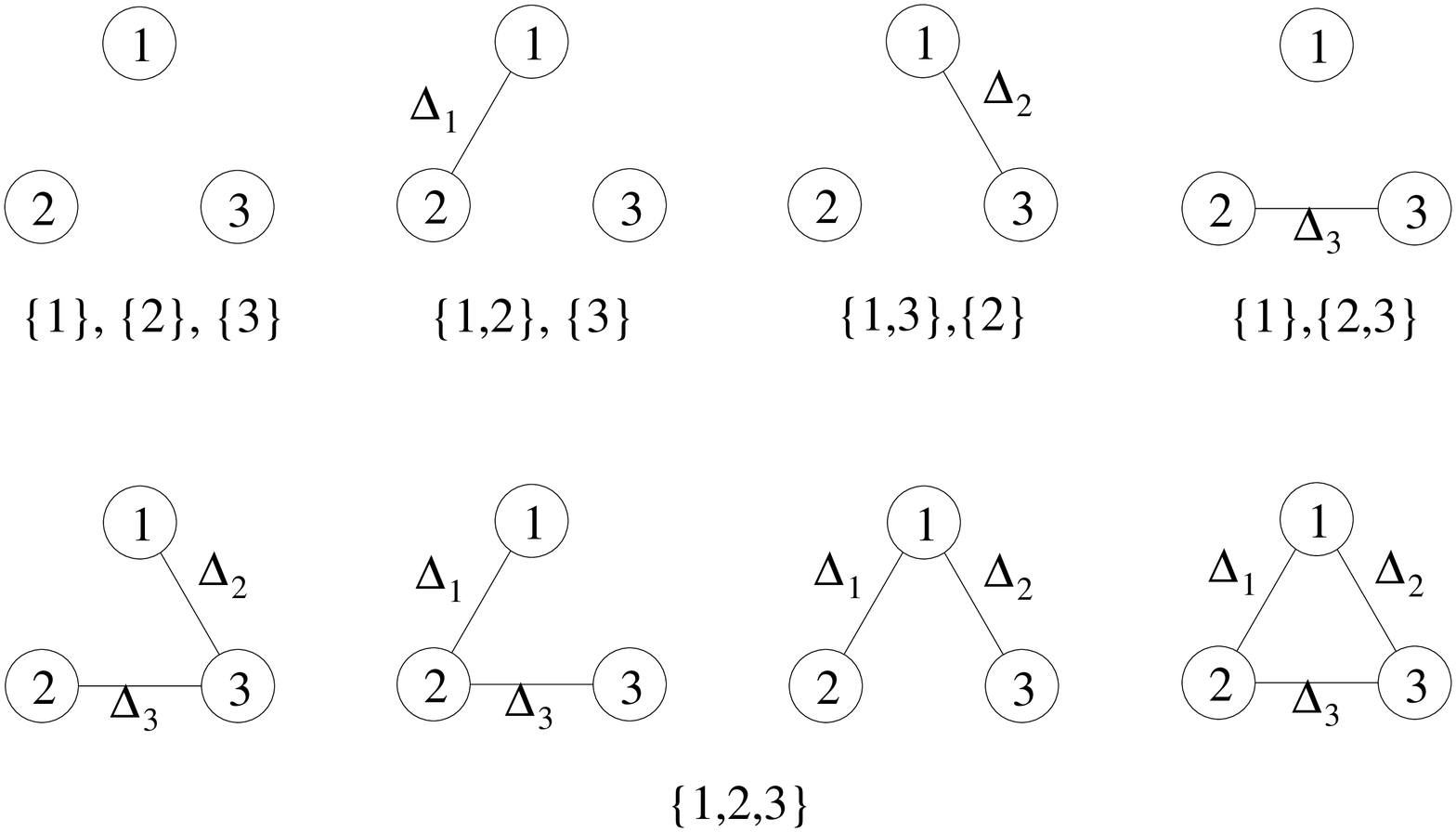}
\vspace{0.5cm}
\caption{Possible patterns of pairing in the toy model with 3 
fermion species, shown as graphs. Nonzero 
gap parameters $\Delta_i$ yield edges in the graphs. Below each graph, 
its components
are listed. The four graphs in the second row have 
identical components and thus are
equivalent in our analysis, even though they would not
have equivalent condensation energies.}
\label{3species}
\end{center}
\end{figure}

Which pairing patterns are possible in this model? 
We consider the gap matrix 
\be \label{gapmatrix}
{\cal M} = \left(\begin{array}{ccc} 
0 & \Delta_1 & \Delta_2 \\ \Delta_1 & 0 & \Delta_3 \\
\Delta_2 & \Delta_3 & 0 \end{array}\right) \,\, ,
\ee
where a nonzero entry 
in the $i$-th row and $j$-th column indicates that fermion $i$ forms Cooper 
pairs with fermion $j$.  The matrix ${\cal M}$ is symmetric because
we are assuming that the Cooper pairs have zero angular momentum,
and are thus antisymmetric in Dirac indices.  We have written
the most general possible gap matrix with diagonal entries set
to zero.  In Section IV, the matrix ${\cal M}$ will be a matrix
in color and flavor space, and we shall only consider pairing
that is antisymmetric in both color and flavor.  This means
that the diagonal entries of ${\cal M}$ will vanish there, so
we make this choice here too.  We will not be solving
a gap equation, and hence will only be interested in which
of the $\Delta$'s are nonzero.  We shall therefore choose
each of $\Delta_1$, $\Delta_2$ and $\Delta_3$ to be either
$\Delta$ or zero, yielding $2^3=8$ possible patterns of pairing.
However, some of these patterns are equivalent in their
consequences for our analysis, as we now explain. We can
denote each of the 8 possible patterns by a graph,
where the vertices of the graph are the
fermions and the edges are the nonzero
gap parameters. In this simple example of three fermion species, 
this might seem notationally
superfluous but in the case of three-flavor quark matter,
with nine fermion species, we shall find that the correspondence between a 
pairing pattern and a graph 
provides a useful tool to classify the multitude of possible pairing patterns.
For further explanation of the elementary
graph-theoretical terms that we shall use in the following, see for instance 
Ref.\ \cite{harary}.  
All 8 graphs are shown in Fig.\ \ref{3species}, and we see that the
last 4 all have the same components --- in this case, actually,
the same single component.  In our analysis, all that matters
is the components since these determine which fermions have 
a common Fermi momenta.  We can therefore consider all the graphs
that have the same components as equivalent, even though
they would have different condensation energies.  This
reduces the number of possible patterns that we must consider,
in this example from 8 to 5.
The number of patterns that we need consider is the number
of possible partitions of the set $\{1,2,3,\ldots,N\}$ into
disjoint, non-empty subsets.  In combinatorics, this number
is called the Bell
number $B_N$, and Fig.\ \ref{3species} shows that
$B_3 = 5$. In general, $B_N$ 
is given by the recursive formula~\cite{combinatorics}  
\be \label{bell}
B_{N} = \sum_{k=0}^{N-1} B_k 
\left(\begin{array}{c} N-1 \\ k \end{array}\right) \,\, ,
\ee 
along with the definition $B_0\equiv 1$.  This yields $B_1=1$, $B_2=2$,
$B_3=5$, $B_4=15$, $\ldots$ $B_9=21147$, $\ldots$.
Returning to our model with $N=3$, we can now discuss
each of the 5 patterns separately, analyzing their neutrality
conditions and looking for stress-free patterns of pairing.
The results are collected in Table \ref{table1}. 

\begin{table} 
\begin{tabular}[t]{|c|c|c|} 
\hline
components   & neutrality & stress-free pairing 
\\ \hline\hline
\;\;\{1\},\{2\},\{3\}\;\; &  $\,\mu_q=\frac{q_3}{q_1^2+q_2^2+q_3^2}\frac{m^2}{2\mu}\,$ &  (no pairing) \\ \hline
\{1,2\},\{3\} & $\mu_q=\frac{1}{3q_3}\frac{m^2}{\mu}$ & 
for any $\mu_q$, if $q_1=q_2$  \\ \hline
\{1,3\},\{2\} & $\mu_q=-\frac{1}{3q_2}\frac{m^2}{2\mu}$ & 
$\mu_q=\frac{1}{q_3-q_1}\frac{m^2}{2\mu}$\\ \hline
\{2,3\},\{1\} & $\mu_q=-\frac{1}{3q_1}\frac{m^2}{2\mu}$ & 
$\mu_q=\frac{1}{q_3-q_2}\frac{m^2}{2\mu}$\\ \hline
\{1,2,3\} & $\mu_q$ arbitrary & 
\,if $q_1=q_2$ and $\mu_q=-\frac{m^2}{6q_1\mu}$\,  
\\ \hline
\end{tabular}
\caption{The neutrality condition (\ref{final}) 
and the condition for stress-free pairing 
(\ref{robust}) for a toy model with 3 fermion species.}
\label{table1}
\end{table}

The first row of this table describes the unpaired scenario. In this case, 
a nonzero $\mu_q$ is 
required, as in the two-species toy model. 

The second, third, and fourth rows of the table describe patterns 
in which two of the fermions form Cooper pairs 
while the third remains unpaired. In the second row, for
example, neutrality is obtained as long as
\be
\mu_q = \frac{q_3}{\frac{1}{2}\left(q_1+q_2\right)^2+q_3^2}\frac{m^2}{2\mu}\ ,
\ee
which yields the result quoted in the table since $q_1+q_2=-q_3$.
The analyses of the neutrality conditions in the third and fourth
rows are similar, and in each of these cases a particular nonzero $\mu_q$
is required.
The conditions for stress-free pairing in the second, third
and fourth rows of the table follow directly from (\ref{robust}) 
and (\ref{chempot}).  For each of the patterns in the second, third
and fourth row of the table, the condition that the 
phase is simultaneously neutral and stress-free turns out to be 
that this is possible only if
$q_1=q_2=-q_3/2$, and that it requires $\mu_q=-m^2/(6q_1\mu)$.

The fifth row of the table describes any of the patterns 
in which
all three fermions pair. (This 
could be any of four different
patterns, as we saw in Fig.~\ref{3species}.)  In this case,
the system is automatically neutral, with any value of $\mu_q$.
However, the pairing is stress-free 
only if $q_1=q_2=-q_3/2$ and $\mu_q=-m^2/(6q_1\mu)$.

Our conclusion for this toy model is that if $q_1\neq q_2$, neutral stress-free
pairing is impossible, whereas if $q_1 = q_2$, we have an 
excess of possibilities, as {\it all} the patterns of pairing 
can be both neutral and stress free.  This latter
result comes about because if 
$q_1=q_2=-q_3/2$ and $\mu_q=-m^2/(6q_1\mu)$ then the unpaired phase
has three equal effective chemical potentials.  The case
of three-flavor quark matter, to which we turn next, is
more analogous to the toy model with $q_1\neq q_2$.

\section{Pairing patterns, color symmetry, and color neutrality} 
\label{realworld}

We now turn to the case of three-flavor quark
matter. The neutrality conditions which we introduced
in previous sections in a general 
and abstract way
shall now be 
those required to ensure that the quark matter
is electrically and color neutral. 
We first discuss possible pairing patterns and the role
of color and flavor symmetries. We then explain how color 
neutrality is taken into account.
This section provides the basis for 
Sec.\ \ref{results}, where we shall 
demonstrate that stress-free pairing is impossible in
three-flavor quark matter.

\subsection{Gap matrix and color/flavor symmetries}

In quark matter with three colors and flavors, 
nine different quarks are present.
Were we to ignore all qualitative features of the
nature of the interaction between quarks in QCD,
we would be forced to consider a general $9 \times 9$
gap matrix ${\cal M}$. Restricting our attention to the
$J=0$ channel which is antisymmetric in Dirac indices (because
$J=1$ pairing yields much smaller gaps~\cite{Alford:1997zt,Alford:2002rz})
makes ${\cal M}$ a symmetric matrix 
specified by 45 independent gap parameters
which, for our purposes, 
are either zero or nonzero. This would lead to $2^{45}$ 
different gap matrices. 
However, as before many of these
gap matrices would have the same set of components, and
the number of patterns that have to be treated separately 
would be given by the Bell number
$B_9 = 21147$, much smaller but still unwieldy.   

In QCD the one-gluon exchange interaction between
two quarks is attractive
in channels which are antisymmetric under
the exchange of the color of the quarks. We know that this is
the channel in which the dominant attraction
between quarks is to be found also by analyzing
the nonperturbative instanton-induced interaction
between quarks, as well as by the qualitative
argument that only by combining two color $[{\bf 3}]_c$ quarks
antisymetrically into a color $[\bar{\bf 3}]_c$ diquark
do we reduce the color flux.   
Cooper pairs that are antisymmetric in color and Dirac indices
must also be antisymmetric in flavor indices, meaning that
we need only consider the antisymmetric flavor 
channel $[\bar{\bf 3}]_f$.
Therefore, we only consider those
$9 \times 9$ gap matrices ${\cal M}$ in 
the direct product space
$[\bar{\bf 3}]_c \otimes [\bar{\bf 3}]_f$, which can be written as
\be \label{Mdefine}
{\cal M} = \D_{ij} \,J_i \otimes I_j \,\, ,
\ee
where  the antisymmetric $3\times 3$ matrices $(J_i)_{jk} = -i\e_{ijk}$ and $(I_i)_{jk} = -i\e_{ijk}$ 
are bases of $[\bar{\bf 3}]_c$ and $[\bar{\bf 3}]_f$, respectively. 
We see that the $9 \times 9$ matrix is fully specified by only 
9 (not 45) gap parameters, namely the 9 entries in 
$\D_{ij}$. Explicitly,
\begin{widetext}
\be \label{gapmatrix9}
{\cal M}=\left(\begin{array}{ccccccccc} 0&0&0&0&-\D_{33}&\D_{32}&0&\D_{23}&-\D_{22}\\
0&0&0&\D_{33}&0&-\D_{31}&-\D_{23}&0&\D_{21}\\ 0&0&0&-\D_{32}&\D_{31}&0&\D_{22}&-\D_{21}&0\\
0&\D_{33}&-\D_{32}&0&0&0&0&-\D_{13}&\D_{12}\\  -\D_{33}&0&\D_{31}&0&0&0&\D_{13}&0&-\D_{11}\\
\D_{32}&-\D_{31}&0&0&0&0&-\D_{12}&\D_{11}&0\\  0&-\D_{23}&\D_{22}&0&\D_{13}&-\D_{12}&0&0&0 \\
\D_{23}&0&-\D_{21}&-\D_{13}&0&\D_{11}&0&0&0\\  -\D_{22}&\D_{21}&0&\D_{12}&-\D_{11}&0&0&0&0
\end{array}\right) \,\, ,
\ee
\end{widetext}
with the quarks ordered $ru$, $rd$, $rs$, $gu$, $gd$, $gs$, $bu$, $bd$, $bs$.
The assumptions we have made, which use only
gross qualitative features of QCD, have
excluded pairing of the same quarks, e.g., $ru$ with $ru$,
and also pairing of quarks of the same color, e.g., $ru$ with $rd$, 
or with the same flavor, e.g., $ru$ with $bu$. 
This is the meaning of the zeros in the matrix ${\cal M}$.
A pairing pattern now is uniquely defined by the (complex)
$3\times 3$ matrix $\Delta_{ij}$. Therefore, within 
our method, the number of possible patterns is
$2^9=512$, corresponding to choosing each of the nine
entries in $\Delta_{ij}$ to be either zero
or $\Delta$.  One of these, with all entries in $\D_{ij}$
set to zero, describes unpaired quark matter. The other 511
describe patterns of conventional BCS color superconducting 
pairing that is antisymmetric in both color and flavor.
Note that some of these patterns may describe more than
one phase.  For example, both the CFL and gCFL phases would
be described here as having a $\Delta_{ij}$ with nonzero
diagonal entries and zero off-diagonal entries, the distinction
between whether the three diagonal entries are equal or unequal
being unnecessary to the determination of whether the pattern
of pairing is stress-free.
In our search for stress-free pairing
described in Section V it will turn out to be most
convenient to consider all 511 patterns even
though, as we explain now, not all are physically distinct.

In QCD color symmetry is unbroken, but flavor symmetry is
explicitly broken both by quark masses and by electric
charge assignments.  Let us first consider, however,
how our analysis would simplify if both flavor and color
symmetries were unbroken.
In this circumstance,
any two matrices ${\cal M}_1$, ${\cal M}_2$ that
are connected by a rotation in color-flavor space
describe the same phase. A color-flavor rotation
$U\otimes V$ where $U\in SU(3)_c$ and $V\in SU(3)_f$ acts on the gap matrix as
\be \label{transformM}
(U\otimes V)\,{\cal M} = \Delta_{ij}\,U\,J_i\,U^T\otimes V\,I_j\,V^T \,\, .
\ee
This rotation can equivalently be applied to the matrix $\Delta$ which transforms as
\be \label{transform}
(U\otimes V)\,\Delta = U^T\,\Delta\, V \,\, , 
\ee
Hence, two
matrices $\Delta_1$ and $\Delta_2$ describe the same phase 
if there exists a rotation $U\otimes V$ such that
$\Delta_1 = U^T\,\Delta_2\, V$. However, for any complex matrix $\Delta_2$ 
there is a choice of unitary matrices $U$ and $V$ such that
$U^T\,\Delta_2\, V$ is diagonal~\cite{strang}.
If flavor and color were
both unbroken symmetries we would therefore be able to restrict our
attention to diagonal matrices $\Delta$, specified by only 3 gap parameters
$\D_{11}$, $\D_{22}$ and $\D_{33}$. The CFL phase corresponds to the
case where these three gap parameters are 
equal~\cite{Alford:1998mk,reviews}.  The gap
equations for the more general 
case where all three can be nonzero and unequal 
have been derived and solved in 
Refs.~\cite{Alford:2003fq,Alford:2004hz,Fukushima:2004zq}, 
which means
that if this were the {\it most} general case the present analysis
would not be necessary.

We must, however, consider the consequences of the 
explicit breaking of flavor symmetry. 
In the case $m_u = m_d = 0$, $m_s\neq 0$, which
we shall consider, there appears to be a remaining unbroken $SU(2)_f$ 
symmetry. 
However, because the $u$ and $d$ quarks have different electric
charges, they will have different effective chemical potentials
in the presence of a nonzero chemical potential $\mu_e$ as is
in general required to impose electric neutrality given the
explicit flavor symmetry breaking introduced by
$m_s\neq 0$.
We therefore assume that there is no residual $SU(2)_f$ symmetry. 
Then, two phases $\Delta_1$ and $\Delta_2$ are equivalent only 
if they are connected by a color rotation,
i.e., if there exists a $U\in SU(3)_c$ with $\Delta_1 = U^T\,\Delta_2$. 
This means that we must consider both diagonal and
non-diagonal matrices $\Delta$. As we explain in the next
subsection, this complicates the analysis of the consequences
of color neutrality. 

As an aside, note that a different analysis is required
for the cases of superfluid $^3$He and for $J=1$ one-flavor
color superconductors, where the relevant symmetry groups
are
$SO(3)_L\times SO(3)_S$ and $SU(3)_c\times SO(3)_J$,
respectively. None of these symmetries are explicitly
broken, and yet in
both cases, non-diagonal gap matrices must be considered, 
as for example in the $A$ phase in both 
systems~\cite{vollhardt,QCDAphase}. 
The reason is that although a general complex matrix
can be diagonalized by left- and right- multiplication
by unitary matrices, it cannot be diagonalized if one
or other matrix is required to be orthogonal rather than unitary. 
In Appendix \ref{symmetries}, we explicitly show
that there is no ``$A$ phase'' in the case where
the unbroken symmetry is $SU(3)_c\times SU(3)_f$.
In three-flavor QCD, unlike in the other
two cases, non-diagonal matrices need only be considered
in the presence of explicit symmetry breaking.

\subsection{Imposing color neutrality}

In order to impose color neutrality within our approach, 
we need to know the corresponding
``color charge chemical potentials'' $\mu_{Q_k}$, 
introduced in Eq.\ (\ref{sumchem}). 
Two such chemical potentials are those originating
from nonzero expectation values of the zeroth component
of the gluon fields $A^0_3$ and $A^0_8$.  These chemical
potentials (equivalently, gluon fields) couple to the
differences between color-charge densities, and so
it is no surprise that in QCD their field equations are such
as to enforce that they take on values corresponding
to the color neutrality conditions 
(\ref{neutral})~\cite{Gerhold:2003js,Dietrich:2003nu}.
However, we show in Appendix~B that,
consistent with the results of 
Ref.~\cite{Buballa:2005bv}, in the case of pairing patterns described by
non-diagonal matrices $\Delta$, color neutrality
may require
nonzero expectation values for the zeroth components of {\it any}
of the gluon fields $A^0_a$, $a=1,\ldots 8$. 
We shall denote these by $\mu_a$, and refer to them
as color chemical potentials, even though this terminology
is only appropriate for $\mu_3$ and $\mu_8$.
All of them except $\mu_3$ and $\mu_8$ appear multiplying
off-diagonal Gell-Mann matrices in color space,
meaning that they cannot be described as chemical potentials
for suitable combinations of color charge densities.
This in turn means that if any except $\mu_3$ or $\mu_8$ are
nonzero, the analysis that we have set up in Section~ II is inapplicable.
And,
we have seen above that because flavor symmetry is explicitly broken
in QCD, we {\it must} consider 
non-diagonal matrices $\Delta$.  It would seem that we are in
trouble.  

If we were trying to derive and solve gap equations for the
patterns of pairing that we analyze, we would not be able
to evade the introduction of non-diagonal color ``chemical 
potentials''.   However, given our more limited goals, 
we can proceed as follows.
Let $\Delta$ be an arbitrary gap matrix, in particular a non-diagonal
one. For this gap matrix, the values of the color chemical 
potentials $\mu_a$ with $a=1,\ldots 8$ required
for color-neutrality can be determined
explicitly as described in Appendix~\ref{tadpoles}.
We set the $\mu_a$'s to these values, and now have
a color-neutral state.  We can now apply any color rotation, 
after which we will still have a color neutral
state.
So, we make that
color rotation $U$ that diagonalizes the matrix 
$\mu_aT_a$, where the $T_a$ are the Gell-Mann matrices.
Since $\mu_aT_a$ is Hermitian, such a transformation 
always exists. Moreover, since $\mu_aT_a$ is
traceless, the diagonalized matrix is also
traceless and thus can be written as a linear combination
of $T_3$ and $T_8$,
\be \label{diag}
U\,\mu_aT_a\,U^\dag = \mu_3^\prime\,T_3 + \mu_8^\prime \, T_8 \,\, ,
\ee
where the new chemical potentials are functions of the old ones, $\mu_3^\prime = 
\mu_3^\prime(\mu_1,\ldots,\mu_8)$, $\mu_8^\prime = \mu_8^\prime(\mu_1,\ldots,\mu_8)$.
The transformation $U$ defines a tranformation for all Gell-Mann matrices, $T_a \to U\,T_a\,U^\dag$,
as well as for the gap matrix, ${\cal M}\to (U\otimes {\bf 1})\,{\cal M}$
or, equivalently, $\Delta\to U^T\Delta$. 
Of course, all the new quantities 
depend on the
chemical potentials $\mu_1,\ldots, \mu_8$. 
As Eq.\ (\ref{invariance}) makes
explicit, if the state was color neutral before this color
rotation it is color neutral after it, assuming of course
that the gap matrix is rotated by the same color rotation $U$
as that applied to the chemical potentials.
This means that 
for an arbitrary $\Delta$ we have found a 
physically equivalent order parameter $U^T\Delta$ for which 
color neutrality requires only nonzero $\mu_3$ and $\mu_8$.

A more formal way of explaining what we have just done is
to note that the
Cartan subalgebra of $SU(3)_c$ has dimension 2, meaning that 
the maximal number 
of mutually commuting generators of $SU(3)_c$ is 2, and thus only two color charges
can be measured simultaneously. This means that 
a phase with arbitrary pairing pattern $\Delta$ 
can be neutralized by the specification of only two linear
combinations of the $\mu_a$'s. In general, these two
combinations $\mu_3'$ and $\mu_8'$ are non-diagonal, involving
all the $\mu_a$'s.  However, by some color rotation $U$
we find a new
basis in which the two chemical potentials that must be nonzero
are just $\mu_3$ and $\mu_8$.   Making this color rotation
changes the pairing pattern from $\Delta$ to $U^T \Delta$.
In general, neither $\Delta$ nor $U^T\Delta$ will be diagonal.

If we wanted to derive and solve gap equations, this
result would be of little practical use as
in order to 
find the ``simpler'' state $U^T\Delta$ 
explicitly one would have to have
solved the whole problem for $\Delta$. 
Our goals are more modest, however. We shall
analyze the most general pairing pattern $\Delta$, in the presence
of only two color chemical potentials
$\mu_3$ and $\mu_8$, and show that stress-free pairing
is impossible.  If the $\Delta$ you find interesting is one
of those which requires other $\mu_a$'s to be nonzero, we leave
it to you to construct $U^T \Delta$.  Since we will have
analyzed all possible $\Delta$'s, your $U^T \Delta$ will
be equal to another $\Delta$ that we have analyzed. Thus,
our analysis below suffices.

\section{Stressed pairing is unavoidable}
\label{results}

In this Section, we 
prove that for any possible pairing pattern in three-flavor quark matter, there
will be a mismatch in the effective chemical potentials of 
at least two fermion species
that form Cooper pairs with each other. 
We end the section with a discussion of 
some specific patterns of pairing
in order to illustrate our method and the result by
example. 

\subsection{General result} \label{method}

\begin{figure} [ht]
\begin{center}
\includegraphics[width=8cm]{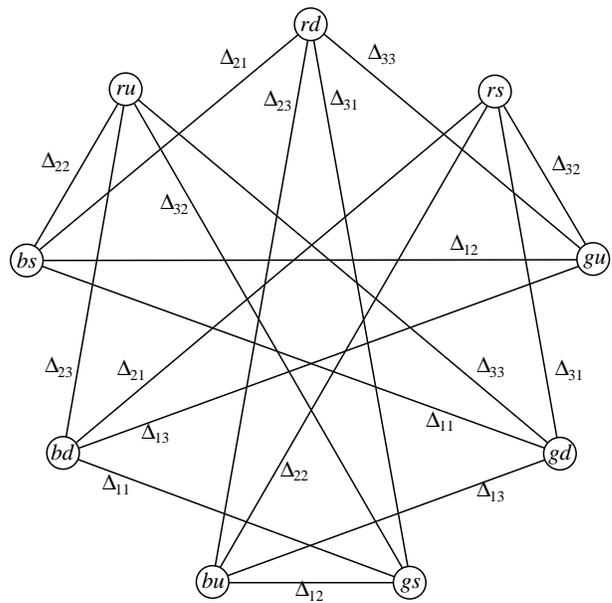}
\vspace{0.5cm}
\caption{Generic graph for possible pairing patterns. The fermions are the vertices while an edge
between two vertices means that the corresponding fermions form Cooper pairs. 
A particular gap matrix
switches off particular edges.}
\label{graph}
\end{center}
\end{figure}

Here we shall apply the general discussion of Section II to the case
of interest, namely nine fermions with effective chemical potentials
as defined in Eq.~(\ref{mueffdefn}) given by 
\bea
\mu_{ru}^{\rm eff} &=& \mu - \frac{2}{3}\mu_e + \frac{1}{2}\mu_3 + \frac{1}{3}\mu_8 \,\, , \non
\mu_{rd}^{\rm eff} &=& \mu + \frac{1}{3}\mu_e + \frac{1}{2}\mu_3 + \frac{1}{3}\mu_8 \,\, , \non
\mu_{rs}^{\rm eff} &=& \mu + \frac{1}{3}\mu_e + \frac{1}{2}\mu_3 + \frac{1}{3}\mu_8 - \frac{m_s^2}{2\mu}\,\, , \non
\mu_{gu}^{\rm eff} &=& \mu - \frac{2}{3}\mu_e - \frac{1}{2}\mu_3 + \frac{1}{3}\mu_8 \,\, , \non
\mu_{gd}^{\rm eff} &=& \mu + \frac{1}{3}\mu_e - \frac{1}{2}\mu_3 + \frac{1}{3}\mu_8 \,\, , \non
\mu_{gs}^{\rm eff} &=& \mu + \frac{1}{3}\mu_e - \frac{1}{2}\mu_3 + \frac{1}{3}\mu_8 - \frac{m_s^2}{2\mu} \,\, , \non
\mu_{bu}^{\rm eff} &=& \mu - \frac{2}{3}\mu_e - \frac{2}{3}\mu_8 \,\, , \non
\mu_{bd}^{\rm eff} &=& \mu + \frac{1}{3}\mu_e - \frac{2}{3}\mu_8 \,\, , \non
\mu_{bs}^{\rm eff} &=& \mu + \frac{1}{3}\mu_e - \frac{2}{3}\mu_8 - \frac{m_s^2}{2\mu}\,\, . 
\label{chemreal}
\eea
Electric neutrality 
is ensured via the electric chemical potential $\mu_e$, while color neutrality is ensured via the
color chemical potentials $\mu_3$ and $\mu_8$, where we have
normalized the corresponding Gell-Mann matrices as $T_3={\rm diag}(1/2,-1/2,0)$
and $T_8={\rm diag}(1/3,1/3,-2/3)$. 
Because 
${\rm Tr}\,Q = {\rm Tr}\,T_3 = {\rm Tr}\,T_8 = 0$,
the condition (\ref{traceless}) is satisfied. We consider color
superconducting phases described by the gap matrix 
(\ref{gapmatrix9}), specified by the $3\times 3$ matrix
$\Delta$ according to (\ref{Mdefine}). 
We may now proceed along the same lines as for 
the toy models in Sec.\ \ref{toy}. 
The only difference 
is the larger number of fermion species
and the resulting larger number
of possible gap matrices. Because our gap matrix is specified
by the nine parameters of $\Delta$, and because we are
only concerned about which parameters are zero and which
nonzero, we have $2^9=512$ patterns of pairing to consider.  
In order to handle this large number of patterns, 
we make use of a one-to-one correspondence between 
a gap matrix (more precisely, between the pattern of
zero and nonzero entries in a gap matrix) and a graph. We
can then
use {\it Maple} routines that know 
the basic terms of graph theory and which, in particular, 
enable us to determine the components of
a graph. This allows us to write a program
that performs the following steps: 
\begin{enumerate}
\item Create a list of all 512 pairing patterns
by setting each entry of $\Delta$ either to 0 
or 1. 

\item Translate each of these gap matrices into a graph where the vertices correspond to the nine
fermions and the edges to nonzero entries of $\Delta$. The generic graph (with all possible 
edges switched on) is shown in Fig.\ \ref{graph}. 

\item Determine the components of each graph. This 
creates a list of 512 sets of component$\{{\cal C}_1,
\ldots, {\cal C}_{c_i}\}$, $i=1,\ldots,512$, one set for
each graph and gap matrix.   Determining the components
of a graph is equivalent to block-diagonalizing the $9\times 9$
matrix ${\cal M}$ of (\ref{gapmatrix}). For example, in the
familiar CFL pattern in which only $\D_{11}$, $\D_{22}$
and $\D_{33}$ are nonzero the components are 
${\cal C}_1=\{ru,gd,bs\}$, 
${\cal C}_2=\{rd,gu\}$, ${\cal C}_3=\{bu,rs\}$, 
and ${\cal C}_4=\{gs,bd\}$.

\item 
There are many instances where several among the 512 graphs 
have the same set of components. In fact, we find only 149 different
sets of components. The
question of whether
a pattern of 
pairing is stress-free or not depends only on its set of components, 
so we need to analyze only 149 cases rather than 512.
One of the sets of components 
contains only single fermions, i.e., it 
corresponds to unpaired quark matter. 
Thus, there are 148 different sets of components corresponding to 
the 511 patterns of color superconducting pairing.

\item For each of these sets of components, 
compute the common Fermi momenta $\nu_1, \ldots, \nu_{c_i}$, 
$i=1,\ldots,149$, according to 
Eq.\ (\ref{common}), and assign them to the respective fermions. Here we use
Eqs.\ (\ref{chemreal}), meaning that
every common Fermi momentum is now a function 
of $\mu$, $\mu_e$, $\mu_3$, $\mu_8$, and $m_s$.

\item  For each set of components, search for a simultaneous solution $(\mu_e,\mu_3,\mu_8)$
of the neutrality equations (\ref{final}) and the equations for stress-free pairing (\ref{robust}).

\end{enumerate}

The result is: {\em For no set of components 
is there a triple $(\mu_e,\mu_3,\mu_8)$ that fulfills Eqs.\ 
(\ref{final}) and (\ref{robust}) simultaneously.} 
(Except for the one that corresponds to unpaired quark matter. In this case, 
both equations are fulfilled
trivially by $(\mu_e,\mu_3,\mu_8)=(m_s^2/(4\mu),0,0)$.)    In all 511
possible patterns of BCS pairing (described by 148 different
sets of components) if the neutrality conditions~(\ref{final}) 
are satisfied there are at least two fermions which pair but
whose values of $\mu_{\rm eff}$ differ by of order $m_s^2/\mu$.  
In all 511 patterns, pairing is stressed, exacting a free energy cost of
order $(m_s^2/\mu)^2\mu^2=m_s^4$. 
Neutral stress-free BCS pairing is impossible.
This is the central conclusion of our analysis.

Before considering examples, let us review how the analysis
of color rotations in Section IV comes into play. 
Remember that 
for any $\Delta$ there is a -- generally complicated and $\mu_a$ dependent --
color rotated $U^T \Delta$ for which color neutrality requires
only nonzero $\mu_3$ and $\mu_8$. 
Consequently, in order to discuss all possible phases it is 
sufficient to discuss all transformed matrices $U^T\Delta$. 
Any possible $U^T\Delta$, however, is included in our description, 
because all we have to know about this
gap matrix is the corresponding set of components, and 
we have included all possible sets of components.
Hence, stress-free pairing is impossible for any conventional
color superconductor.

In order to explain this argument somewhat differently, 
consider a specific set of components. 
This set may originate from many different gap matrices. 
It will be impossible to describe
some of these gap matrices by the ansatz for the color chemical potentials
in Eqs.\ (\ref{chemreal}), because they require
nonzero $\mu_a$'s other than $\mu_3$ and $\mu_8$. 
Perhaps even all possible gap matrices corresponding to this 
specific set of components
are described incorrectly. 
This would have been a problem for our method if we had found cases
in which Eqs.\ (\ref{final}) and (\ref{robust}) {\it were} 
simultaneously fulfilled. Then, no direct
conclusion would have been possible because we would not 
have know whether there was even one phase ``hidden'' behind
the set of components which allows for stress-free pairing that {\it is} 
correctly described using only 
the color chemical potentials $\mu_3$ and $\mu_8$. 
Fortunately, our result (no stress-free pairing)
is rigorous because the argument goes the other way around: 
{\em any} pattern of pairing is hidden
behind one of our set of components, and 
since no set of components exhibits stress-free pairing we are able
to reach a general conclusion.

\begin{table*}[th]
\begin{tabular}{||c||c|c|c||c||c|c|c||} 
\hline\hline& & & & & & &\\
$\;\;$diagonal order parameters$\;\;$ & $\mu_e$ & $\mu_3$ & $\mu_8$ & 
$\;\;$non-diagonal order parameters$\;\;$ & $\mu_e$ & $\mu_3$ & $\mu_8$\\ & & & & & & &\\ \hline\hline & & & & & & &\\
$\Delta_1=\left(\begin{array}{ccc}1 & 0 & 0 \\ 0&0&0 \\ 0&0&0\end{array}\right)$ & $\frac{m_s^2}{4\mu}$ &0&0&   
$\Delta_8=\left(\begin{array}{ccc}1 & 1 & 0 \\ 0&0&0 \\ 0&0&0\end{array}\right)$ & $\frac{m_s^2}{4\mu}$ &0&0 \\  
& & & & & & & \\ \hline& & & & & & &\\
$\Delta_2=\left(\begin{array}{ccc}0 & 0 & 0 \\ 0&1&0 \\ 0&0&0\end{array}\right)$ & 0&0&0&   
$\Delta_9=\left(\begin{array}{ccc}1 & 0 & 1 \\ 0&0&0 \\ 0&0&0\end{array}\right)$ & $\frac{m_s^2}{4\mu}$ &0&0 \\  
& & & & & & &\\ \hline& & & & & & &\\
$\Delta_3=\left(\begin{array}{ccc}0 & 0 & 0 \\ 0&0&0 \\ 0&0&1\end{array}\right)$ & $\frac{m_s^2}{2\mu}$ &0&0&   
$\Delta_{10}=\left(\begin{array}{ccc}0 & 1 & 1 \\ 0&0&0 \\ 0&0&0\end{array}\right)$ & $\frac{m_s^2}{4\mu}$ &0&0 \\  
& & & & & & &\\ \hline& & & & & & &\\
$\Delta_4=\left(\begin{array}{ccc}1 & 0 & 0 \\ 0&1&0 \\ 0&0&0\end{array}\right)$ & 0&0&$-\frac{m_s^2}{2\mu}$&   
$\Delta_{11}=\left(\begin{array}{ccc}1 & 1 & 0 \\ 0&0&1 \\ 0&0&0\end{array}\right)$ & $\mu_e$&0&
$-\mu_e+\frac{m_s^2}{\mu}$ \\  
& & & & & & &\\ \hline& & & & & & &\\
$\Delta_5=\left(\begin{array}{ccc}1 & 0 & 0 \\ 0&0&0 \\ 0&0&1\end{array}\right)$ & $\frac{m_s^2}{2\mu}$ &
 $\frac{m_s^2}{2\mu}$ &$-\frac{m_s^2}{4\mu}$ &   
$\Delta_{12}=\left(\begin{array}{ccc}1 & 0 & 1 \\ 0&1&0 \\ 0&0&0\end{array}\right)$ & $\mu_e$&0&
$-\mu_e-\frac{m_s^2}{2\mu}$ \\  
& & & & & & &\\ \hline& & & & & & &\\
$\Delta_6=\left(\begin{array}{ccc}0 & 0 & 0 \\ 0&1&0 \\ 0&0&1\end{array}\right)$ & $\mu_e$&$\mu_e-\frac{m_s^2}{3\mu}$ &
 $\frac{\mu_e}{2}$ &   
$\Delta_{13}=\left(\begin{array}{ccc}0 & 1 & 1 \\ 1&0&0 \\ 0&0&0\end{array}\right)$ & $\mu_e$&0&
$2\mu_e-\frac{m_s^2}{2\mu}$ \\  
& & & & & & &\\ \hline& & & & & & &\\
$\Delta_7=\left(\begin{array}{ccc}1 & 0 & 0 \\ 0&1&0 \\ 0&0&1\end{array}\right)$ &$\;\;$ $\mu_e$$\;\;$&$\;\;$$\mu_e$$\;\;$&
$\;\;$$\frac{\mu_e}{2}-\frac{m_s^2}{2\mu}$$\;\;$&   
$\Delta_{14}=\left(\begin{array}{ccc}1 & 1 & 1 \\ 0&0&0 \\ 0&0&0\end{array}\right)$ & $\;\;$$\frac{m_s^2}{4\mu}$$\;\;$ &
$\;\;$$-\frac{2}{3}\mu_8$$\;\;$ &$\;\;$$\mu_8$$\;\;$   
 \\ & & & & & & &\\ \hline\hline   
\end{tabular}
\caption{Order parameters for which $\mu_1=\mu_2=\mu_4=\ldots =\mu_7=0$.
The electric and 
color chemical potentials satisfying the neutrality constraints are
given for each pattern of pairing. In those cases in which the neutrality
conditions leave one linear combination of $\mu_e$, $\mu_3$ and $\mu_8$
undetermined, we write the chemical potentials in terms of
an unspecified $\mu_e$ (or $\mu_8$ in the one case in which
the undetermined combination does not involve $\mu_e$.) }
\label{tablephases}
\end{table*}

\subsection{Specific examples} \label{examples}

We will illustrate our result using as examples
gap matrices $\Delta$ for which we know
that color neutrality can be 
imposed be adjusting only $\mu_3$ and $\mu_8$, leaving
$\mu_1=\mu_2=\mu_4=\ldots =\mu_7=0$. 
According to the results of Appendix \ref{evaluate},
we  may consider all matrices which have at most
one nonzero entry per column. 
By choosing the entries
of $\Delta$ either zero or one, this gives rise to 64 gap matrices 
(each column may either 
be completely zero or may have a 1 at three possible entries, 
hence there are $4^3=64$ options).
As we prove in Appendix~\ref{phases},
14 of these gap matrices are independent while
all the others can be obtained from one of these 14 via
a color rotation.
In Table \ref{tablephases} we list these 14 matrices. 

\begin{figure*} [ht]
\begin{center}
\hbox{\includegraphics[width=0.45\textwidth]{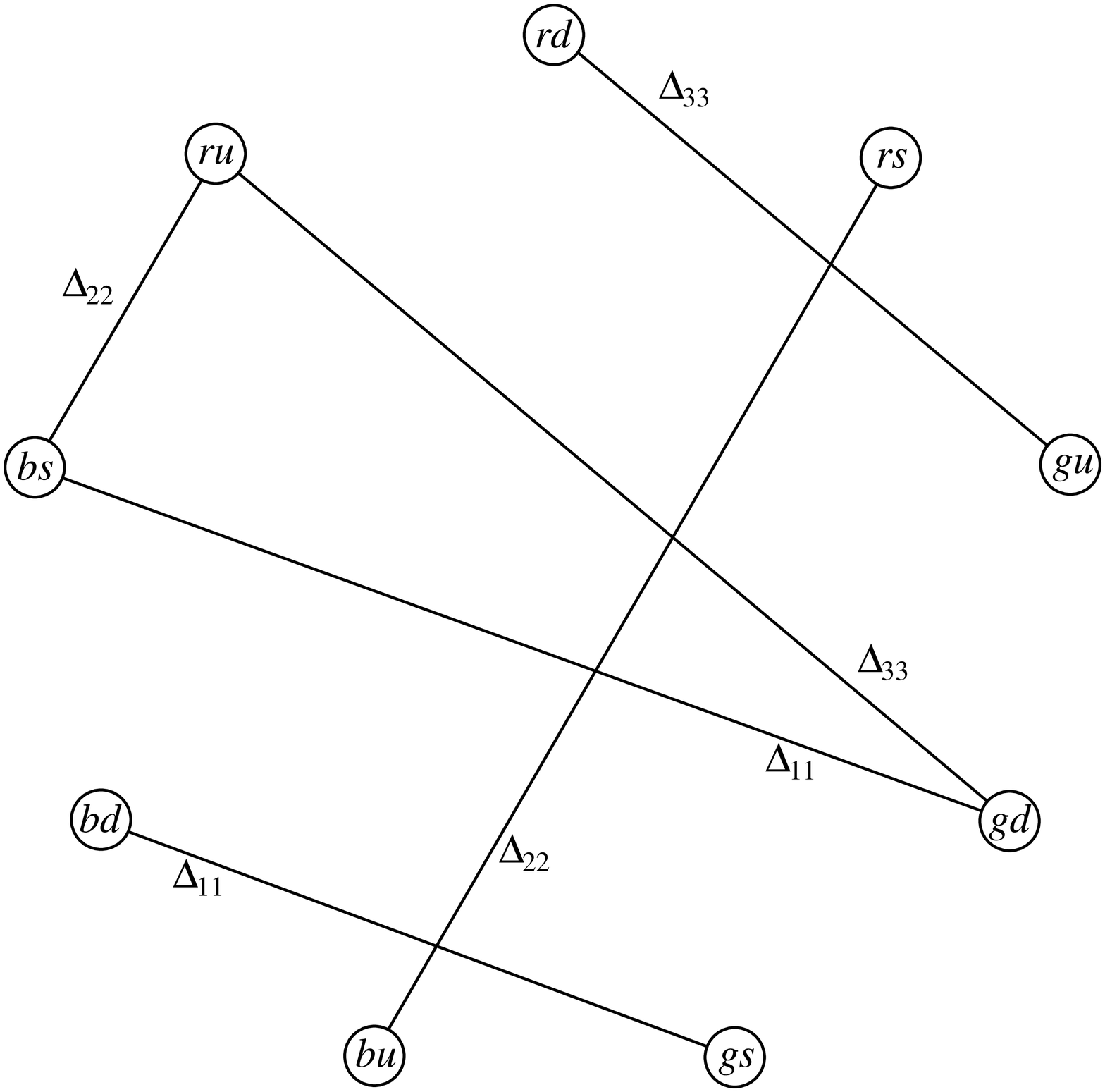}
\hspace{1.5cm}\includegraphics[width=0.45\textwidth]{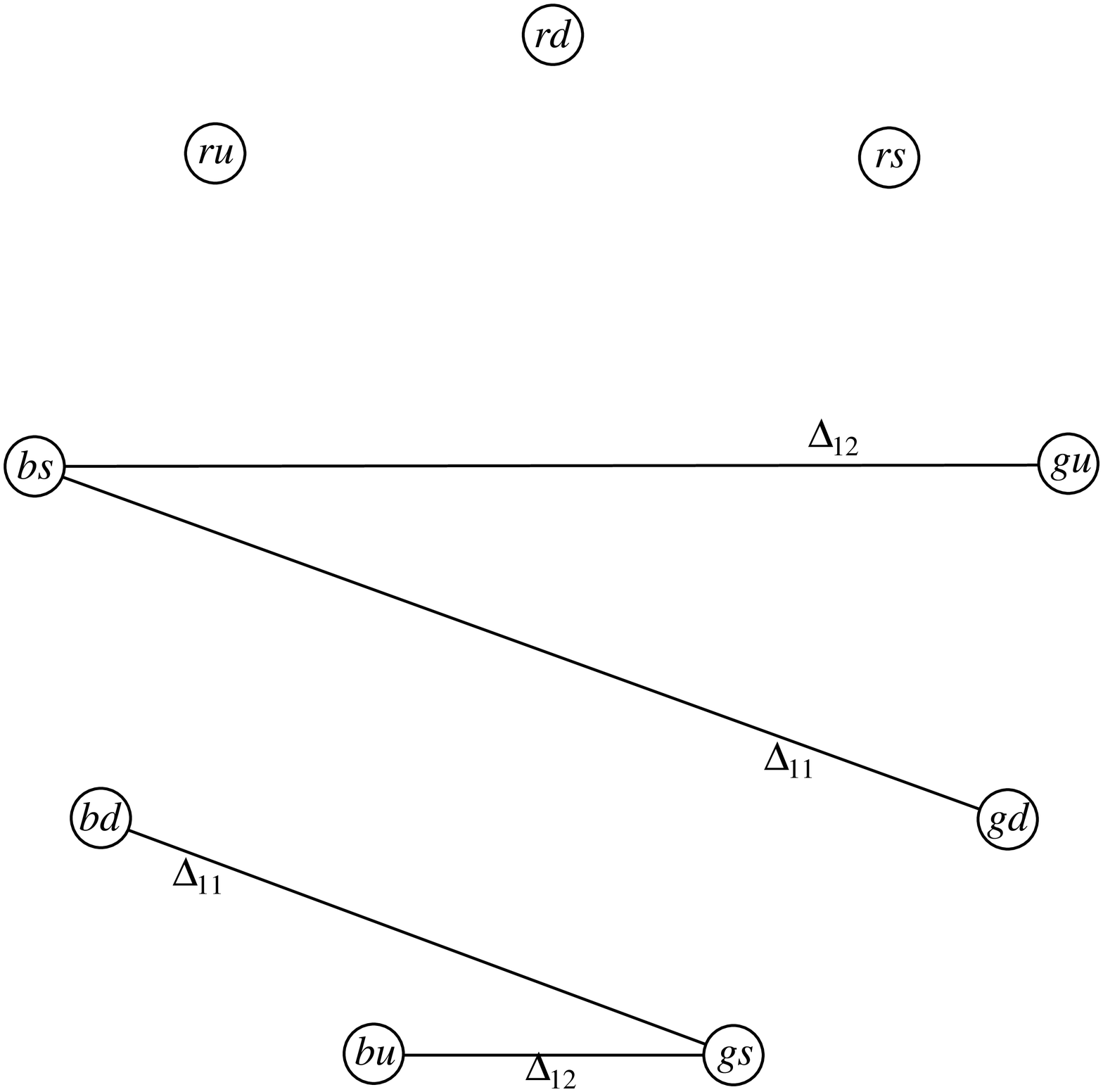}}
\vspace{0.5cm}
\caption{Graphs for the patterns of pairing 
$\Delta_7$ (CFL) and $\Delta_8$, cf.\ Table \ref{tablephases}.}
\label{figgraph2}
\end{center}
\end{figure*}

The left half 
of Table \ref{tablephases} contains 
all diagonal order parameters. One of them, here
denoted by $\Delta_7$, is the pattern of
pairing for the CFL phase. 
The corresponding graph is shown in the left panel of
Fig.\ \ref{figgraph2}. The graphs of all
the other diagonal order parameters in Table II are subgraphs
of that for $\D_7$ shown in Fig.\ \ref{figgraph2}, i.e., 
they are obtained by removing one or more edges of this graph. 
The left half of the table includes several phases that are familiar
from the literature. For example, the $\D_3$ pattern of pairing
is the 2SC phase~\cite{Alford:1997zt,Rapp:1997zu}, the uSC and dSC phases
described in Refs.~\cite{Iida:2003cc,Fukushima:2004zq} 
are described by the patterns $\D_6$
and $\D_5$, and the $\D_2$ pattern
is the 2SCus phase~\cite{Alford:2002kj,Alford:2004hz}. 
The right half of Table~\ref{tablephases}
shows the seven patterns of pairing 
among our 14 examples which have non-diagonal gap
matrices. If flavor symmetry were unbroken,
each of these patterns would be equivalent to one of the 
diagonal ones. As flavor symmetry is
explicitly broken, however,
these patterns of pairing are physically distinct from the ones in the
left half of the table. 
The electric and color chemical potentials in the table were obtained
from the neutrality conditions (\ref{final}).

Let us discuss two of the patterns of pairing in more
detail, explicitly showing 
that the results for the chemical potentials in Table \ref{tablephases}
do not allow for stress-free pairing. 
We do so for the patterns $\Delta_7$ (CFL) and $\Delta_{8}$. 
The corresponding graphs are shown in Fig.\ \ref{figgraph2}.
We take the explicit results from the table for 
$\mu_e$, $\mu_3$, and $\mu_8$ and use them to evaluate 
the effective chemical potentials of Eqs.\ (\ref{chemreal}). 
For the pattern of pairing 
$\Delta_{8}$, we find 
(showing only the components with more than a single quark)
\be
{\cal C}_1: \quad \begin{array}{l} 
\mu^{\rm eff}_{gu} = \mu - \frac{m_s^2}{6\mu} \\
\mu^{\rm eff}_{gd} = \mu + \frac{m_s^2}{12\mu} \\
\mu^{\rm eff}_{bs} = \mu - \frac{5m_s^2}{12\mu} \end{array} \,\, ,
\qquad 
{\cal C}_2: \quad \begin{array}{l} 
\mu^{\rm eff}_{gs} = \mu - \frac{5m_s^2}{12\mu} \\
\mu^{\rm eff}_{bu} = \mu - \frac{m_s^2}{6\mu} \\
\mu^{\rm eff}_{bd} = \mu + \frac{m_s^2}{12\mu} \end{array} \,\, .
\ee
It is obvious that this state does not exhibit stress-free pairing.
For the CFL pattern $\Delta_7$, we obtain
\begin{widetext}
\bea
&&{\cal C}_1: \quad \mu^{\rm eff}_{ru} = \mu^{\rm eff}_{gd} = \mu^{\rm eff}_{bs} = \mu-\frac{m_s^2}{6\mu} \,\, , \qquad {\cal C}_2: \quad
\begin{array}{l} \mu^{\rm eff}_{rd} = \mu + \mu_e - \frac{m_s^2}{6\mu} \\
                 \mu^{\rm eff}_{gu} = \mu - \mu_e - \frac{m_s^2}{6\mu} \end{array}\,\, , \non
&&{\cal C}_3: \quad \begin{array}{l}\mu^{\rm eff}_{rs} = \mu - \frac{2m_s^2}{3\mu} + \mu_e 
\\ \mu^{\rm eff}_{bu} = \mu + \frac{m_s^2}{3\mu} - \mu_e \end{array} \,\, , 
\qquad {\cal C}_4:\quad 
\begin{array}{l}\mu^{\rm eff}_{gs} = \mu - \frac{2m_s^2}{3\mu} \\ 
\mu^{\rm eff}_{bd} = \mu + \frac{m_s^2}{3\mu} \end{array} \,\, .
\eea
\end{widetext}
Even though one linear combination of $\mu_e$, $\mu_3$ and
$\mu_8$ is not fixed by the neutrality conditions, reflected
here by the unspecified value of $\mu_e$,
we see that we can nevertheless
draw a firm conclusion.
We see that the fourth component,
${\cal C}_4$, exhibits a mismatch of 
effective chemical potentials for any value of $\mu_e$,
meaning that stress-free pairing is impossible.
(In a treatment that goes
beyond the order to which we are working, 
the electron chemical 
potential can be determined by including 
the electron pressure into the free energy. 
The result in the CFL case is 
$\mu_e=0$~\cite{Rajagopal:2000ff,Alford:2002kj}, 
leading to the values for the effective chemical
potentials illustrated in Fig.\ \ref{figsplit}.) 
There are many patterns of pairing among the 511 that we have analyzed
for which, at least at the order at  which
we are working, one combination of the chemical potentials is not
fixed by the neutrality conditions. In all such cases we have
shown that stress-free pairing cannot arise for any value
of the undetermined combination of chemical potentials.

\section{Conclusion} \label{conclusion}

We have investigated all possible patterns of 
conventional BCS pairing in color-superconducting
three-flavor quark matter in which the Cooper pairs
are antisymmetric in color and flavor.  
We have analyzed 511 patterns of pairing, but it should
be noted that many of them correspond to more than one phase, since
our analysis only keeps track of which entries in the gap matrix
are zero and which nonzero, not whether different nonzero entries are
equal or unequal. Also, pushing 
in the opposite direction, not all of the 511 
patterns are distinct: some are related by color rotation.  Regardless,
our analysis applies to every possible pattern, only a handful
of which have been studied before.

At asymptotic densities,
where the differences among the quark masses can be neglected, 
previous work has shown that the CFL phase is 
favored~\cite{reviews}.  As the
density is decreased, CFL pairing requires that the system pay
a free energy cost that is parametrically of order 
$(m_s^2/\mu)^2\mu^2 = m_s^4$,
in order to gain a free energy benefit of order $\D^2 \mu^2$.
This guarantees that below some $\mu$ of order
$m_s^2/\D$, the CFL phase must break down.  In this
investigation we have shown that  {\it all} 511 possible pairing
patterns suffer the same fate.  All of them exact a 
free energy cost of order $m_s^4$; none are stress-free.

There are, then, three classes of
possible answers to the question of
what succeeds the CFL phase as a function
of decreasing density.   First, if $m_s^2/\D$ is small enough the CFL
phase will be succeeded directly by nuclear matter: 
the quark matter $\rightarrow$ nuclear matter transition will
occur before $\mu$ drops to the critical value of order $m_s^2/\D$
at which CFL pairing breaks down.  
The second possibility is that the
CFL phase is succeeded by a color-superconducting state 
with unconventional pairing, not simply as in BCS theory. Such
a state could
be one which is spatially inhomogeneous, one which
includes quasiparticles which are paired but gapless, 
one which includes counterpropagating currents, 
one which includes meson condensates, or one which
involves any combination of these possibilities.
Third, it is a logical possibility
that the CFL phase is succeeded by some among the 510 less symmetrically
paired BCS phases
that we have considered, for example the 2SC phase.  However, the conclusion
of our work is that any of these must also break down below a $\mu$
of order $m_s^2/\D$, meaning that any of these must also be succeeded
by some state with non-BCS pairing.  None of the 511 possible
patterns of BCS pairing can survive in a regime in 
which $\mu\ll m_s^2/\D$.

\begin{acknowledgments}

We thank E.\ Gubankova, H.\ Malekzadeh, M.\ Mannarelli, D.\ Rischke, R.\ Sharma, and I.\ Shovkovy
for valuable comments and discussions. 
AS acknowledges financial support from the German Academic Exchange 
Service (DAAD). 
KR acknowledges the hospitality of the Nuclear Theory Group at LBNL.
This research was supported in part by the Director, Office of
Science, Office of
Nuclear Physics,  of the U.S.\
Department of Energy under contract \#DE-AC02-05CH11231
and cooperative research agreement
\#DF-FC02-94ER40818.

\end{acknowledgments}

\appendix 

\section{Why there is no ``{\it A} phase'' in the case of a $SU(3)_c\times SU(3)_f$ symmetry}
\label{symmetries}

In this Appendix, we elaborate on the difference between a three-flavor color superconductor on the one hand and
superfluid $^3$He and a one-flavor color superconductor on the other hand. All three systems are described 
by an order parameter
parametrized by 
a $3\times 3$ matrix $\Delta$. The reason is that in all cases the 
Cooper pairing occurs in channels which are triplets with respect
to two different symmetries.
In a three-flavor color superconductor these are
the color and flavor (anti)triplets, in superfluid $^3$He the 
angular momentum and spin triplets, and 
in a one-flavor color superconductor the color and total 
angular momentum (anti)triplets.
In the latter two systems there is a non-diagonal order parameter matrix
\be
\Delta_A = \frac{1}{\sqrt{2}}\left(\begin{array}{ccc} 0&0&0\\0&0&0\\1&i&0\end{array}\right) \,\, ,
\ee
describing the so-called {\it A} phase \cite{vollhardt,QCDAphase}. 
This phase is genuinely different from all 
phases that can be described by a diagonal matrix $\Delta$ because there is no symmetry rotation that tranforms
$\Delta_A$ into a diagonal matrix. In the case of $^3$He, the 
symmetry rotations are elements of the
group $SO(3)_L\times SO(3)_S$, where $SO(3)_L$ and $SO(3)_S$ are the angular momentum and spin 
groups, respectively. This group acts on $\Delta$ as given 
in Eq.\ (\ref{transform}). Suppose there
is a transformation $U\otimes V$ with $U\in SO(3)_L$ and $V\in SO(3)_S$ that diagonalizes $\Delta_A$. 
This requires that the off-diagonal 
elements of the resulting matrix vanish,
\be
\sum_{k\ell} u_{ki}\,\Delta_{k\ell}\,v_{\ell j} =0 \qquad \mbox{for}\,\, i\neq j \,\, ,
\ee
where $u_{ki}$ and $v_{\ell j}$ are the entries of the matrices $U$ and $V$ respectively.
Or, explicitly,
\begin{subequations}
\bea
u_{31}(v_{12}+iv_{22})&=&u_{31}(v_{13}+iv_{23})=0 \,\, , \\
u_{32}(v_{11}+iv_{21})&=&u_{32}(v_{13}+iv_{23})=0 \,\, , \\
u_{33}(v_{11}+iv_{21})&=&u_{33}(v_{12}+iv_{22})=0 \,\, .
\eea
\end{subequations}
Since $U$ is unitary and thus invertible, at least one of the entries $u_{31}$, $u_{32}$, and 
$u_{33}$ must be nonzero. Without loss of generality, we may assume that $u_{31}\neq 0$. Since all
$v_{ij}$'s are real, this means that $v_{12}=v_{22}=v_{13}=v_{23}=0$. Therefore,
\be
V\,V^T =\left(\begin{array}{ccc} v_{11}^2 & v_{21}v_{11} & v_{31}v_{11} \\ v_{21}v_{11} & v_{21}^2 &
v_{31}v_{21} \\ v_{31}v_{11} & v_{31}v_{21} & v_{31}^2 + v_{32}^2 + v_{33}^2\end{array}\right) \,\, .
\ee
The orthogonality of $V$ then implies $v_{11}^2=v_{21}^2=1$ and $v_{21}v_{11}=0$. These two 
conditions contradict each other. Consequently, there is no transformation $U\otimes V$ that diagonalizes
$\Delta_A$.

Since we only needed 
the $v_{ij}$'s to be real, the same proof holds for complex $u_{ij}$'s, i.e., for
the symmetry group $SU(3)_c\times SO(3)_J$, relevant for a one-flavor color superconductor, where 
$J$ is the total angular momentum. 

However, in the case of the symmetry group $SU(3)_c\times SU(3)_f$, hence for unitary transformations
$U$ and $V$, any matrix $\Delta$ can be diagonalized
by $U^T\Delta V$, see for example Ref.\ \cite{strang}. 
In particular, a potential ``{\it A} phase'' order parameter
is diagonalized by the pure flavor rotation
\be
U= {\bf 1} \,\, , \qquad 
V = \frac{1}{\sqrt{2}}\left(\begin{array}{ccc} 0&i&1\\0&-1&-i\\ \sqrt{2}&0&0\end{array}\right) \,\, ,
\ee
where $U\in SU(3)_c$ and $V\in SU(3)_f$.
With these matrices we find
\be
U^T\,\Delta_A\,V = \left(\begin{array}{ccc} 0&0&0\\0&0&0\\0&0&1\end{array}\right)  \,\, .
\ee  
Hence, if $SU(3)_c\times SU(3)_f$ were unbroken in three-flavor QCD,
the ``{\it A} phase'' would be equivalent to one with a diagonal $\D$.  

As described in Section IV and manifest in Section V, however, the
explicit breaking of $SU(3)_f$ in QCD by a nonzero strange quark
mass means that a generic nondiagonal $\D$ is {\it not} equivalent
to a diagonal $\D$.  We must therefore consider all 511 patterns
of pairing including those which are analogues of the $A$ phase.

\section{Tadpoles and color neutrality}
\label{tadpoles}

In this Appendix we determine which color chemical
potentials have to be nonzero to neutralize a three-flavor
color superconducting phase whose gap matrix ${\cal M}$
is specified by (\ref{Mdefine}), with $\Delta$ an arbitrary $3\times 3$ matrix.
In QCD, these 
chemical potentials originate from 
nonzero expectation values of the corresponding zeroth component
of the gluon fields $A^0_a$, $a=1,\ldots 8$. 
In the mean-field approximation, 
the Yang-Mills equation for the gluon fields requires 
the zeroth component of the 
quark tadpole contributions to vanish 
\cite{Gerhold:2003js,Dietrich:2003nu},
\be \label{tadvanish}
{\cal T}_a^0 = 0 \,\, , \qquad a=1,\ldots ,8 \,\, .
\ee
Here, ``quark tadpole'' refers
to the Feynman diagram of a quark loop with a gluon leg. 
The explicit expressions are given by \cite{Dietrich:2003nu}
\be \label{tad}
{\cal T}_a^\mu = -\frac{g}{2}{\rm Tr}[\Gamma_a^\mu\,{\cal S}] \,\, , 
\ee
where $g$ is the strong coupling constant. 
Before discussing the explicit structure of the quark
propagator ${\cal S}$, let us point out the physical 
meaning of the eight equations (\ref{tadvanish}).
The two tadpoles ${\cal T}_3$ and ${\cal T}_8$ are
proportional to color density differences, and
even though the other tadpoles cannot be interpreted
in this way this makes clear that satisfying the requirement
that all tadpoles vanish results in a color-neutral state.
One often replaces the precise statement that the eight
$\mu_a$'s take on values such that the  eight ${\cal T}_a$'s
vanish by the loose statement that color neutrality
is achieved by adjusting the color chemical potentials
$\mu_a$ such that all color densities vanish.  As
explained in Section IV, this loose statement can always
be rendered precise by a color rotation, which also rotates $\Delta$.

In normal quark matter, the tadpoles vanish even with vanishing gluon fields $A^0_1 = \ldots = A^0_8=0$.  
In superconducting quark matter, however, 
nonvanishing gluon 
fields have to be introduced in order to fulfil Eqs.\ (\ref{tadvanish}). 
Which gluon fields attain a nonzero expectation value depends
on the pairing pattern.

\subsection{Formalism}


In Eq.\ (\ref{tad}), the vertex 
$\Gamma_a^\mu\equiv {\rm diag}(\g^\mu T_a,-\g^\mu T_a^T)$, where $T_a$ are the Gell-Mann matrices,
and the quark propagator 
\be
{\cal S} = \left(\begin{array}{cc} G^+ & \Xi^- \\ \Xi^+ & G^- \end{array}\right)  
\ee
are given in the Nambu-Gorkov basis. We do not need to specify the exact form
of the anomalous propagators $\Xi^{\pm}$ because they drop out in the expression for the tadpole after
taking trace over Nambu-Gorkov space. The diagonal components $G^{\pm}$ are the propagators for quarks and
charge-conjugated quarks,
\be \label{quarkprop}
G^{\pm} \equiv \{[G_0^{\pm}]^{-1}  - \Phi^{\mp}\,G_0^{\mp}\,\Phi^{\pm}\}^{-1} \,\, ,
\ee
where $G_0^{\pm}$ is the tree-level quark (charge-conjugate quark) propagator and $\Phi^{\mp}$ 
the off-diagonal components of the quark self-energy. In general, in order to ensure color
neutrality, all 8 color chemical potentials enter the (inverse) tree-level propagators,
\be \label{treeprop}
[G_0^{\pm}]^{-1} = \gamma^\mu K_\mu \pm \g_0 \,(\mu+\mu_a \,T_a + \mu_e\,Q) - m \,\, ,
\ee
where summation over $a=1,\ldots,8$ is implied, and $K_\mu$ 
is the quark four-momentum. The matrices
$Q$ and $m$ are proportional to the identity in color space,
and are 
diagonal in flavor space. They 
account for the quark electric charges and masses,
respectively.
For the following qualitative discussion, we are not interested in the
exact value of the tadpoles. 
We are rather interested in the effect of the color-flavor structure on ${\cal T}_a^0$. 
Thus, we may drop all Dirac and Nambu-Gorkov structure and write the tadpoles as
\be \label{tadpole}
{\cal T}_a^0 = {\rm Tr}\left\{T_a \,\left[F + \mu_bT_b - {\cal M}^\dagger\,(F - \mu_bT_b)^{-1}\,{\cal M}
\right]^{-1}\right\} \,\, ,
\ee
where $F$ is a diagonal matrix in flavor space. We have replaced the (inverse) propagator by the schematic
expression $[G_0^{\pm}]^{-1} \to F\pm \mu_aT_a$ and $\Phi^+ \to {\cal M}$, $\Phi^-\to {\cal M}^\dag$. 

We remark that the tadpoles remain invariant
under color rotations of the Gell-Mann matrices and the gap matrix,
\be \label{invariance}
{\cal T}_a^0\,[T_a,{\cal M}] = {\cal T}_a^0\,[U\,T_a\,U^\dag,U\,{\cal M}] \,\, .
\ee
Although this is not difficult to see, we present here an explicit
proof.  
The proof is basically a repeated application of
the formula
\be
(F+U\,M\,U^\dag)^{-1} = U\,(F+M)^{-1}\,U^\dag \,\, ,
\ee
which is valid for unitary $U$ and $U\,F\,U^\dag = F$. In the 
following, $F$ is a matrix in flavor
space, i.e., $F={\bf 1}\otimes F$, and $U$ acts on color space, i.e., 
$U\,F\,U^\dag = U\,U^\dag\otimes F = F$;
$M$ can be a matrix in both color and flavor spaces, 
say $M=m_{ij}A_i\otimes B_j$ with complex coefficients
$m_{ij}$ and color and flavor matrices $A_i$ and $B_j$, respectively. 
Then, by making use of Eq.\ (\ref{transformM}),
\begin{widetext}
\bea
{\cal T}_a^0\,[U\,T_a\,U^\dag,U\,{\cal M}] &=& 
{\rm Tr}\left\{UT_aU^\dag\,\left[F + \mu_b\,UT_bU^\dag - \Delta_{ij}^*\Delta_{mn}\,UJ_iU^\dag\,I_j
(F - \mu_b\,UT_bU^\dag)^{-1}\,UJ_mU^\dag\,I_n\right]^{-1}\right\} \non
&=& {\rm Tr}\left\{UT_aU^\dag\,\left[F + \mu_b\,UT_bU^\dag - \Delta_{ij}^*\Delta_{mn}\,UJ_i\,I_j
(F - \mu_b\,T_b)^{-1}\,J_mU^\dag\,I_n\right]^{-1}\right\} \non
&=& {\rm Tr}\left\{UT_aU^\dag\,\left[F + U\,\{\mu_bT_b - {\cal M}^\dag(F - \mu_b\,T_b)^{-1}{\cal M}\}
\,U^\dag \,\right]^{-1}\right\} \non
&=& {\rm Tr}\left\{T_a\,\left[F + \mu_b T_b - {\cal M}^\dag(F - \mu_b\,T_b)^{-1}{\cal M}
\right]^{-1}\right\} \non
&=& {\cal T}_a^0\,[T_a,{\cal M}] \,\, ,
\eea
\end{widetext}
which is Eq.\ (\ref{invariance}).

\subsection{Evaluating tadpoles}
\label{evaluate}

We may now use the general form of the tadpoles (\ref{tadpole}) 
to investigate which chemical potentials are needed
for a given gap matrix $\Delta$.
To this end, we have 
to evaluate the tadpoles for
{\em vanishing} color chemical potentials $\mu_1=\ldots = \mu_8=0$. 
Let us denote these quantities by ${\cal T}_a^0(0)$. 
{}From Eq.\ (\ref{tadpole}) we obtain
\be \label{rigorous}
{\cal T}_a^0(0) \equiv {\rm Tr}\left[T_a \,\left(F - {\cal M}^\dagger\,F^{-1}\,{\cal M}\right)^{-1}\right] 
\,\, ,
\ee
where $F\equiv {\rm diag}(f_1,f_2,f_3)$ accounts 
for the explicit flavor symmetry breaking.
A nonvanishing tadpole, ${\cal T}_a^0(0)\neq 0$, indicates 
that the corresponding color chemical potential $\mu_a$ has to be nonzero
in order to obtain a color neutral state. 
The converse, namely the statement that a vanishing
tadpole, ${\cal T}_a^0(0)= 0$, indicates that the 
state can be color-neutralized
with the corresponding color chemical potential 
$\mu_a$ set to zero, is not immediate. We
have not found a general argument that
precludes the concern that
after introducing a nonzero color chemical 
$\mu_a$ that cancels the initially nonzero ${\cal T}_a^0(0)$,  
some other tadpole ${\cal T}_b^0$ could 
now be nonvanishing, even though 
${\cal T}_b^0(0)$ was zero. 
We have, however, checked that for very many pairing
patterns, including all those in Table \ref{tablephases}
and many more, this concern does not arise.  We have checked
this for patterns with multiple nonzero entries per column
in $\Delta$, with complex entries in $\Delta$, and for patterns
with up to five different nonzero ${\cal T}_a^0(0)$'s, which
of course means nonzero ${\cal T}_a^0(0)$'s that do not commute.
This suggests the following conjecture: 
First, use Eq.\ (\ref{rigorous}) to 
compute the tadpoles with all the $\mu_a$'s set
to zero, namely the ${\cal T}_a^0(0)$ for all $a=1,\ldots 8$.
Now, keep $\mu_a=0$ only for those values of $a$ for which
${\cal T}_a^0(0)=0$; for the other values of $a$, reintroduce
$\mu_a$ as a free variable.  
Then, use  Eq.\ (\ref{tadpole}) to compute all eight
tadpoles ${\cal T}_b^0$ as a function of the nonzero $\mu_a$'s.   
We conjecture that, for all pairing patterns $\Delta$, the 
${\cal T}_b^0$ computed in this way will vanish
for all values of $b$ for which ${\cal T}_b^0(0)$ was
found to vanish in the first step.
We do not have a proof of this conjecture, but we  know of no
counterexample and, as described above, we have checked
many examples. 

Having described the consequences of which $T_a^0(0)$ are zero
and which nonzero, 
we turn now to the evaluation of the expression (\ref{rigorous}). 
We begin with the simplest possible situation, setting 
$F={\bf 1}$, i.e. switching off all flavor symmetry breaking 
terms. We are left with the inversion of the matrix ${\bf 1} - {\cal M}^\dagger {\cal M}$. It can be formally 
inverted by writing the Hermitian matrix 
${\cal M}^\dagger {\cal M}$ in terms of its eigenvalues $\lambda_i$ and the projectors 
onto the corresponding eigenspaces ${\cal P}_i$,
\be
{\cal M}^\dagger {\cal M} = \sum_{i=1}^n \lambda_i\, {\cal P}_i \,\, , 
\ee
where we denoted the number of different eigenvalues by $n$. Then, since $\{{\cal P}_1,\ldots,{\cal P}_n\}$
is a complete set of orthogonal projectors,
\be
({\bf 1} - {\cal M}^\dagger {\cal M})^{-1} = \sum_{i=1}^n \frac{1}{1-\lambda_i}\, {\cal P}_i \,\, .
\ee
The projectors are given by (see for example Appendix A of Ref.\ \cite{thesis})
\be
{\cal P}_i = \prod_{j\neq i}^n\frac{{\cal M}^\dagger {\cal M} - \lambda_j}{\lambda_i - \lambda_j} 
\,\, .
\ee
As a further simplification, let us assume that ${\cal M}^\dagger {\cal M}$ has at most two different 
eigenvalues. Then, the projectors ${\cal P}_i$ are a linear combination of the matrices ${\bf 1}$
and ${\cal M}^\dagger {\cal M}$ (in the case of more than two different eigenvalues, higher powers of 
${\cal M}^\dagger {\cal M}$ are present). Since ${\rm Tr}\,T_a=0$, we obtain
\be \label{tadpolesymmetric}
{\cal T}_a^0(0) \propto {\rm Tr}[T_a\,{\cal M}^\dagger {\cal M}] = 
-2\,{\rm Tr}[T_a\,\Delta\,\Delta^\dagger] \,\, , 
\ee
where Eq.\ (\ref{Mdefine}) as well as ${\rm Tr}[T_aJ_iJ_j] = -(T_a)_{ij}$ and ${\rm Tr}[I_iI_j] = 2\delta_{ij}$
have been used. This simple form of the tadpole can be applied for example to the CFL phase with $m_s=0$. In this
case $\Delta = {\bf 1}$ and the eigenvalues of ${\cal M}^\dagger {\cal M}$ are 4 and 1. 
We immediately conclude ${\cal T}_a^0(0)=0$ for all $a$. This reflects
the fact that in the absence of flavor symmetry breaking
the CFL phase is automatically 
color neutral, i.e., there is no nonzero color chemical potential.

The simplest way to introduce explicit flavor symmetry breaking ``by hand'' is to insert a flavor matrix 
$F$ into Eq.\ (\ref{tadpolesymmetric}),
\be \label{tadpolenonsymmetric}
{\cal T}_a^0(0) \propto {\rm Tr}[T_a\,{\cal M}^\dagger\,F\, {\cal M}] = 
-\,{\rm Tr}[T_a\,\Delta\,F^\prime\Delta^\dagger] \,\, , 
\ee
where we used ${\rm Tr}[I_i\,F\,I_j] = \delta_{ij}\,{\rm Tr}\,F - F_{ij}$ and defined 
$F^\prime\equiv {\rm Tr}\,[F]\cdot{\bf 1} - F$. 
(Since the only purpose of $F^\prime$ is to break the 
flavor symmetry, we may drop the prime in the following.) The expressions 
for the tadpoles can now be evaluated explicitly:
\begin{widetext}
\begin{subequations} \label{alltadpoles}
\bea 
{\cal T}_1^0(0) &\propto& \;\;\; f_1\,{\rm Re}[\D_{11}\D_{21}^*] + f_2\,\,{\rm Re}[\D_{12}\D_{22}^*] 
+ f_3\,{\rm Re}[\D_{13}\D_{23}^*] \\
{\cal T}_2^0(0) &\propto& i\left(f_1\,{\rm Im}[\D_{11}\D_{21}^*] + 
f_2\,{\rm Im}[\D_{12}\D_{22}^*] + f_3\,{\rm Im}[\D_{13}\D_{23}^*]\right)  \\
{\cal T}_3^0(0) &\propto& \frac{1}{2}[f_1(|\D_{11}|^2 - |\D_{21}|^2) + f_2(|\D_{12}|^2 
-|\D_{22}|^2) + f_3(|\D_{13}|^2 - |\D_{23}|^2)] \\
{\cal T}_4^0(0) &\propto& \;\;\;f_1\,{\rm Re}[\D_{11}\D_{31}^*] + f_2\,{\rm Re}[\D_{12}\D_{32}^*] 
+ f_3\,{\rm Re}[\D_{13}\D_{33}^*] \\
{\cal T}_5^0(0) &\propto& i\left(f_1\,{\rm Im}[\D_{11}\D_{31}^*] + 
f_2\,{\rm Im}[\D_{12}\D_{32}^*] + f_3\,{\rm Im}[\D_{13}\D_{33}^*]\right) \\
{\cal T}_6^0(0) &\propto& \;\;\;f_1\,{\rm Re}[\D_{21}\D_{31}^*] + f_2\,{\rm Re}[\D_{22}\D_{32}^*] 
+ f_3\,{\rm Re}[\D_{23}\D_{33}^*]  \\
{\cal T}_7^0(0) &\propto&  i\left(f_1\,{\rm Im}[\D_{21}\D_{31}^*] + 
f_2\,{\rm Im}[\D_{22}\D_{32}^*] + f_3\,{\rm Im}[\D_{23}\D_{33}^*]\right) \\
{\cal T}_8^0(0) &\propto& \frac{1}{2\sqrt{3}}[f_1(|\D_{11}|^2 + |\D_{21}|^2 - 2|\D_{31}|^2)
 + f_2(|\D_{12}|^2 + |\D_{22}|^2 - 2|\D_{32}|^2) \non
&& + f_3(|\D_{13}|^2 + |\D_{23}|^2 - 2|\D_{33}|^2)] \,\, .
\eea
\end{subequations}
\end{widetext}
Note that we have obtained Eqs.\ (\ref{alltadpoles}) by inserting flavor-breaking
by hand, and have not been able to derive them from the rigorous
expression (\ref{rigorous}). However, by inserting various gap matrices (including all the examples in
Table \ref{tablephases} and many more) explicitly into Eq. (\ref{rigorous}), we have checked that (\ref{alltadpoles}) and
the rigorous form (\ref{rigorous}) yield the same conclusion. I.e., a given gap
matrix gives rise to the same nonzero tadpoles ${\cal T}_a^0(0)$ whether
we use (\ref{rigorous}) or (\ref{alltadpoles}).
We can also use the two-flavor result given
below in Eqs. (\ref{alltadpoles2}) as another check:  Eqs.\ (\ref{alltadpoles2}) can be obtained
from Eqs.\ (\ref{alltadpoles}) immediately, and we {\it have} been able to
derive (\ref{alltadpoles2}) rigorously.

As an example of the use of (\ref{alltadpoles}), we can now apply these equations to the CFL and 
gCFL phases,
in both of which the only nonzero $\D$'s are $\D_{11}$,
$\D_{22}$ and $\D_{33}$.  From 
this we see immediately that in the CFL and gCFL phases,
the tadpoles ${\cal T}_1^0(0)$, ${\cal T}_2^0(0)$, ${\cal T}_4^0(0),\ldots,{\cal T}_7^0(0)$
vanish whereas the tadpoles ${\cal T}_3^0(0)$ 
and ${\cal T}_8^0(0)$ may be nonzero due
to the explicit flavor symmetry breaking.  In fact, 
in the CFL phase $f_1=f_2$
and $|\D_{11}|=|\D_{22}|$, which makes ${\cal T}_3^0(0)=0$ leaving
only ${\cal T}_8^0(0)$ nonzero. This is
consistent with the result of Refs.~\cite{Alford:2002kj,Steiner:2002gx}
that the CFL phase is color neutral with $\mu_3=0$ and 
$\mu_8=-m_s^2/(2\mu)$. 
In the gCFL phase, however,
$|\D_{11}|\neq|\D_{22}|$ and both ${\cal T}_3^0(0)$ and 
${\cal T}_8^0(0)$ are nonzero.
Consequently, Eqs.\ (\ref{alltadpoles}) 
indicate that in the gapless CFL phase color
neutrality can be ensured via the introduction
of suitable nonzero values of $\mu_3$ and $\mu_8$, 
as was shown explicitly in Refs.~\cite{Alford:2003fq,Alford:2004hz}. 
Furthermore, Eqs.\ (\ref{alltadpoles}) show that $\mu_1$,
$\mu_2$, $\mu_4\ldots \mu_7$ vanish, as assumed 
in Refs.~\cite{Alford:2003fq,Alford:2004hz,Alford:2004nf,Ruster:2004eg,Fukushima:2004zq,Alford:2004zr,Abuki:2004zk,Ruster:2005jc} and demonstrated in Ref.~\cite{Abuki:2005ms}.
     
We can apply Eqs.\ (\ref{alltadpoles}) more generally as follows. 
We note that for $(a,b)$ being either $(1,2)$, $(4,5)$, or $(6,7)$ 
the following statement holds:
both ${\cal T}_a$ and ${\cal T}_b$ are zero if at least one member 
of each of the following three
pairs is zero: $(\D_{m1},\D_{n1})$, $(\D_{m2},\D_{n2})$, $(\D_{m3},\D_{n3})$. Here, $(m,n)=(1,2)$ for
$(a,b)=(1,2)$, $(m,n)=(1,3)$ for $(a,b)=(4,5)$, and $(m,n)=(2,3)$ for $(a,b)=(6,7)$. 
(The argument is
straightforward. Consider one of these pairs, say $(\D_{m1},\D_{n1})$. 
To make both ${\cal T}_a^0(0)$ and ${\cal T}_b^0(0)$
vanish, according to Eqs.\ (\ref{alltadpoles}) we have to require 
${\rm Re}[\D_{m1}\D_{n1}^*] = {\rm Im}[\D_{m1}\D_{n1}^*] = 0$.
Hence $\D_{m1}\D_{n1}^*$ and therefore either $\D_{m1}$ or $\D_{n1}$ (or both) must vanish.)  
Taking the case $(a,b)=(1,2)$ as an example, we have learned that
both ${\cal T}_1^0(0)$ and ${\cal T}_2^0(0)$ vanish if either 
the first or the second (or both)
entries in each column of the order parameter matrix $\Delta$ vanish. 
Including the other two cases, we reach the following conclusion:
{\em ${\cal T}_1^0(0)={\cal T}_2^0(0)=
{\cal T}_4^0(0)=\ldots={\cal T}_7^0(0)=0$ 
if each column of the matrix $\Delta$ contains at most one nonzero entry.}

We can also derive an analogous result for
a two-flavor color superconductor. This case differs from the current one in the structure of the 
order parameter. 
Since condensation takes place in the flavor-singlet channel, the order parameter is given by 
a three-vector $\Delta\equiv(\Delta_1,\Delta_2,\Delta_3)$ than by a $3\times 3$ matrix. 
The gap matrix is ${\cal M} = \Delta_i\,J_i \,\tau_2$, with the
antisymmetric flavor singlet described by the 
second Pauli matrix $\tau_2$.
Repeating the same steps as for the three-flavor case, we find 
${\rm Tr}[T_a\,{\cal M}^\dagger\,F\,{\cal M}] = -{\rm Tr}[F]\,\Delta^\dagger T_a \Delta$, 
with $F\equiv{\rm diag}(f_1,f_2)$, and thus  
\begin{subequations} \label{alltadpoles2}
\bea 
{\cal T}_1^0(0) &\propto& (f_1+f_2)\,{\rm Re}[\D_{1}\D_{2}^*] \\
{\cal T}_2^0(0) &\propto& i(f_1+f_2)\,{\rm Im}[\D_{1}\D_{2}^*]  \\
{\cal T}_3^0(0) &\propto& \frac{1}{2}(f_1+f_2)\,(|\D_{1}|^2 - |\D_{2}|^2) \\
{\cal T}_4^0(0) &\propto& (f_1+f_2)\,{\rm Re}[\D_{1}\D_{3}^*] \\
{\cal T}_5^0(0) &\propto& i(f_1+f_2)\,{\rm Im}[\D_{1}\D_{3}^*] \\
{\cal T}_6^0(0) &\propto& (f_1+f_2)\,{\rm Re}[\D_{2}\D_{3}^*]  \\
{\cal T}_7^0(0) &\propto& i(f_1+f_2)\,{\rm Im}[\D_{2}\D_{3}^*]  \\
{\cal T}_8^0(0) &\propto& \frac{f_1+f_2}{2\sqrt{3}}\,(|\D_{1}|^2 + |\D_{2}|^2 - 2|\D_{3}|^2)\,\, .
\eea
\end{subequations}
We see that 
${\cal T}_1^0(0)={\cal T}_2^0(0)
={\cal T}_4^0(0)=\ldots={\cal T}_7^0(0)=0$ 
if the vector $\Delta$ contains at most one nonzero entry.

The main results of this appendix can be summarized as follows:
\begin{itemize}
\item The tadpoles ${\cal T}_1^0(0)$, ${\cal T}_2^0(0)$, ${\cal T}_4^0(0)$, ${\cal T}_5^0(0)$, ${\cal T}_6^0(0)$,
${\cal T}_7^0(0)$ vanish if each column in the gap matrix $\Delta$ contains at most one nonzero 
entry. (In a 
two-flavor color superconductor, they vanish if the vector $\Delta$ contains at
most one nonzero entry.)
\item
If there is at least one column in the gap matrix $\D$
with more than one nonzero entry,
the ``usual'' color
chemical potentials $\mu_3$ and $\mu_8$ are not 
sufficient to ensure color neutrality. (In the two flavor case, this arises if the vector $\D$ has
more than one nonzero entry.
The possibility
of chemical potentials other than $\mu_3$ and $\mu_8$ in 
a two-flavor color superconductor has been
discussed in detail in Ref.\ \cite{Buballa:2005bv}.)
\end{itemize}
Note that in the two flavor case, a color rotation can
always be found that takes a generic gap vector $\Delta$ 
into the 2SC pattern $\Delta=(0,0,1)$, 
meaning that chemical potentials other than $\mu_3$ and $\mu_8$
can always be avoided.  In the three flavor case, on the other hand, 
a color rotation alone cannot turn a generic
gap matrix $\D$ into one which has only one nonzero entry per column.
We must therefore proceed as described in Sections IV and V, making
a color rotation that diagonalizes $\mu_a T_a$ but does not simplify
$\D$, meaning that we have to analyze an exhaustive list of $\D$'s.

\section{Superconducting phases in Table \ref{tablephases}} \label{phases}

In this Appendix, we prove that 

\begin{enumerate}
\item Any of the 64 patterns of pairing described by
$3\times 3$ matrices $\D$  which have at most one 
nonzero entry per column can be related 
to one of the 14 patterns of pairing  in Table \ref{tablephases} 
by a color rotation.

\item The 14 patterns of pairing in Table \ref{tablephases} are not connected
by any color rotation.
\end{enumerate} 

To prove the first point, we have to show that we obtain all 64 patterns
of pairing that have at most one
nonzero entry per column by color rotations from the phases in Table \ref{tablephases}. One can simply apply a
color permutation to each of the phases.  A color permutation is equivalent to a permutation
of the rows in the matrix $\Delta$.  Therefore, $\Delta_1$ 
in Table II represents 3 matrices, and so do $\Delta_2$ and $\Delta_3$. 
$\Delta_4$ through $\Delta_7$ represent 6 matrices each, etc. 
In total, in the same order as in Table II, 
3+3+3+6+6+6+6+3+3+3+6+6+6+3=63. The 64th is the zero matrix.

To prove the second point, we have to show that for any pair
$(\Delta_i,\Delta_j)$ chosen from among the 14 patterns of
pairing in Table II, 
there is no unitary matrix $U \in SU(3)$ such that $\Delta_j=U\,\Delta_i$. 
This is obvious  
if there is one column that is completely zero in $\Delta_i$ but has a 
nonzero entry in $\Delta_j$. In this case, there is no matrix --- whether 
unitary or not --- that can 
transform $\Delta_i$ into $\Delta_j$. 
Therefore, the 14 matrices can be divided into the following subsets:
$\{\Delta_1\}$, $\{\Delta_2\}$, $\{\Delta_3\}$, $\{\Delta_4,\Delta_8\}$,
$\{\Delta_5,\Delta_9\}$, $\{\Delta_6,\Delta_{10}\}$, 
and $\{\Delta_7,\Delta_{11},\Delta_{12},\Delta_{13},\Delta_{14}\}$.
For any pair $(\Delta_i,\Delta_j)$ with $\Delta_i$ and $\Delta_j$ 
chosen from two different subsets, the 
statement is obviously true. The remaining cases must 
be treated separately. Suppose there is a $U$
such that $U\,\Delta_4=\Delta_8$. It follows that $u_{21}=u_{22}=u_{31}=u_{32}=0$, where $u_{ij}$ are the 
entries of the matrix $U$. Then,
\be
UU^\dag = \left(\begin{array}{ccc}|u_{11}|^2 + |u_{12}|^2 + |u_{13}|^2 & u_{13}u_{23}^* & 
u_{13}u_{33}^* \\ u_{23}u_{13}^* & |u_{23}|^2 & u_{23}u_{33}^* \\ u_{33}u_{13}^* & 
u_{33}u_{23}^* & |u_{33}|^2 \end{array}\right) \,\, .
\ee
The requirement that $U$ be unitary leads to the contradictory 
requirements $|u_{23}|^2 = |u_{33}|^2 = 1$
and $u_{23}u_{33}^* = 0$. Consequently, $\Delta_4$ and $\Delta_8$ are not equivalent. By symmetry,
the same argument applies to the pairs $(\Delta_5,\Delta_9)$ and 
$(\Delta_6,\Delta_{10})$. 
Only the
last of the subsets remains. 
Suppose that there is a $U$ with $U\,\Delta_7=\Delta_{11}$. This leads 
to a vanishing row in $U$, $u_{31} = u_{32} = u_{33} = 0$. Thus, $U$ cannot be invertible and thus not unitary.
With the same argument one shows that $\Delta_7$ is not equivalent to $\Delta_{12}$, $\Delta_{13}$, and 
$\Delta_{14}$. Now suppose that $U\,\Delta_{11}=\Delta_{12}$. Then, there must be a vanishing column
in $U$, $u_{11} = u_{21} = u_{31} = 0$. Hence this $U$ cannot be unitary. By the same argument,
the equivalence of $\Delta_{11}$, $\Delta_{13}$ and $\Delta_{12}$, $\Delta_{13}$
is excluded. For $U\,\Delta_{11}=\Delta_{14}$ we find $u_{21}=u_{22}=u_{31}=u_{32}=0$. As shown above,
this cannot be fulfilled by a unitary matrix. Therefore, $\Delta_{11}$ and $\Delta_{14}$ are not
equivalent, and, by symmetry, the same holds for the pairs 
$(\Delta_{12},\Delta_{14})$ and 
$(\Delta_{13},\Delta_{14})$. This proves that all 14 matrices are independent.

\end{document}